\documentclass[aps,prb,twocolumn,superscriptaddress]{revtex4}
\usepackage{graphicx}
\usepackage{amsmath}
\usepackage{bbold}
\usepackage[colorlinks=true, citecolor=blue, urlcolor=blue, linkcolor=blue,bookmarks=false,hypertexnames=true]{hyperref} 

\usepackage{doi}

\begin{document}
\graphicspath{}

\title{The effect of valley, spin and band nesting on the electronic properties of gated quantum dots in a single layer of transition metal dichalcogenides (TMDCs)}

\author{Maciej Bieniek}
\affiliation{Department of Physics, University of Ottawa, Ottawa, Ontario, Canada K1N 6N5}
\affiliation{Department of Theoretical Physics, Wroc\l aw University of Science and Technology, Wybrze\.ze Wyspia\'nskiego 27, 50-370 Wroc\l aw, Poland}

\author{Ludmi\l a Szulakowska}
\affiliation{Department of Physics, University of Ottawa, Ottawa, Ontario, Canada K1N 6N5}

\author{Pawe\l \ Hawrylak}
\affiliation{Department of Physics, University of Ottawa, Ottawa, Ontario, Canada K1N 6N5}

\date{\today}

\begin{abstract}
We present here results of an atomistic theory of electrons confined by metallic gates in a single layer of transition metal dichalcogenides. The electronic states are described by the tight-binding model and computed using a computational box with periodic boundary conditions including up to millions of atoms. The confinement is modelled with a parabolic confining potential over the computational box. With this methodology applied to MoS$_\textrm{2}$, we find a two-fold degenerate energy spectrum of electrons confined in the two non-equivalent K-valleys by the metallic gates as well as six-fold degenerate spectrum associated with Q-valleys. We compare the electron spectrum with the energy levels of electrons confined in GaAs/GaAlAs and in self-assembled quantum dots. We discuss the role of spin splitting and topological moments on the K and Q valley electronic states in quantum dots with sizes comparable to experiment.
\end{abstract}


\maketitle

\section{Introduction}

There is currently interest in electron spin based qubits \cite{brum_coupled_1997, loss_quantum_1998,tarucha_shellfilling_1996,ciorga_addition_2000, meier_quantum_2003, elzerman_single-shot_2004, hanson_spins_2007,kyriakidis_prb2002} and circuits realized in field-effect transistors (FET) \cite{hsieh_physics_2012, nowack_single-shot_2011, botzem_tuning_2018, west_gate-based_2019, ito_four-single_spin_2018}. Since the first localization of a single electron in a GaAs/GaAlAs FET by metallic gates \cite{ciorga_addition_2000}, circuits in GaAs and silicon have been realized \cite{lim_observation_2009, kouwen_single_2010, mar_bias-controlled_2011, kawakami_electrical_2014, maurand_cmos_2016}. A similar effort was directed towards understanding electronic states of electrons confined in self-assembled quantum dots \cite{bayer_hidden_2000, michler_single_2003, raymond_excitonic_2004, mar_bias-controlled_2011}. In both cases, the single particle spectrum was understood in terms of a spectrum of two harmonic oscillators and directly observed in InAs/GaAs quantum dots \cite{raymond_excitonic_2004}. In these structures electrons are localized in a volume containing millions of atoms, hence nuclear spins and atomic vibrations contribute to decoherence of electron spins. Recent realization of semiconductor layers with atomic thickness \cite{castro_neto_electronic_2009,geim_van_2013, ihn_graphene_2010, mak_atomically_2010,  manzeli_2d_2017,guclu_graphene_2014,scrace_magnetoluminescence_2015} has opened the possibility of confining single electrons to a few atom thick layers, potentially significantly increasing operating temperature and coherence of electron spin qubits. The conduction band minima in both graphene and transition metal dichalcogenides (TMDCs) are localized in two non-equivalent valleys opening the possibility of using the valley degree of freedom as an additional variable \cite{kadantsev_electronic_2012, cao_valley-selective_2012, mak_control_2012, jones_optical_2013, mak_valley_2014, szulakowska_electronic_2019}. The low energy conduction and valence band states in TMDCs can also be approximated by a massive Dirac Fermion Hamiltonian with resulting nontrivial valley and topological properties \cite{rose_spin-_2013, kormanyos_monolayer_2013, szulakowska_electronic_2019}. The potential of massive Dirac Fermions as qubits has been recognized by a number of theoretical \cite{pawlowski_valley_2018, chirolli_strain-induced_2019, kormanyos_spin-orbit_2014,liu_intervalley_2014, dias_robust_2016, wu_spin-valley_2016, brooks_spin-degenerate_2017, qu_tunable_2017, szechenyi_impurity-assisted_2018} and experimental \cite{pisoni_gate-tunable_2018, lu_optical_2019, brotons-gisbert_coulomb_2019,song_gate_2015} works. Much of this interest in TMDCs based qubits is the possibility of manipulating the 'valley' degree of freedom, e.g., with  circularly polarized light \cite{cao_valley-selective_2012, zeng_valley_2012, mak_control_2012}. In addition to the massive Dirac Fermion physics and the two K-valleys, TMDCs exhibit 3 additional minima per valley in the conduction band at Q points. The presence of Q points \cite{kadantsev_electronic_2012, bernardi_extraordinary_2013, szulakowska_electronic_2019} results in the band nesting and strong coupling to light. Even though all TMDCs share a honeycomb crystal structure, direct bandgaps at K and -K valleys, strong excitonic effects and different metal atoms (Mo or W) change the spin ordering and dispersion of conduction bands at K and Q points, allowing for nontrivial spin dependence of confined electrons. Moreover, the electronic properties of TMDCs can be engineered with composition \cite{wang_electronics_2012, cong_enhanced_2015, mu_electronic_2018, miao_tunable_2018}, strain \cite{frisenda_biaxial_2017, chirolli_strain-induced_2019}, substrate \cite{yun_schottky_2016, man_protecting_2016} or external electromagnetic fields \cite{qu_tunable_2011, scrace_magnetoluminescence_2015, lee_valley_2017, wang_valley_2017, chen_magnetic_2018}, facilitating their application in spin- and valley- based electronics.

Recently, quantum dots (QDs) in graphene, bilayer graphene and TMDCs have been realized as either finite size clusters with different edge termination \cite{guttinger_spin_2010, guclu_graphene_2014, mcguire_growth_2016, wang_quantum_2017, wang_electrical_2018, pisoni_gate-tunable_2018} or by electrostatic confinement with lateral metal electrodes \cite{volk_electronic_2011, allen_gate-defined_2012, eich_spin_2018, pisoni_gate-tunable_2018, wang_electrical_2018, kurzmann_ihn_2019}.  QDs are also formed by combining different TMDC crystals in the plane, which form a potential well \cite{huang_lateral_2014}. 

Gate defined quantum dots avoid the need for atomistic control of the edges. Several groups reported on the creation of finite size electron droplets using metallic gates and observed Coulomb blockade in transport \cite{wang_electrical_2018, pisoni_gate-tunable_2018, brotons-gisbert_coulomb_2019}. Gated quantum dots combined with large trion binding energies allowed for optical probing of excitons in TMDC QDs \cite{lu_optical_2019, pisoni_gate-tunable_2018, wang_electrical_2018, brotons-gisbert_coulomb_2019, chakraborty_3d_2018}. Gerardot and co-workers demonstrated single electron and hole transfer into WSe$_\textrm{2}$ QDs \cite{brotons-gisbert_coulomb_2019} and Srivastava and co-workers estimated long valley lifetimes of localized holes in these QDs due to excess charge \cite{lu_optical_2019}. Charged excitons have also been proven to supress valley scattering by Vamivakas and co-workers \cite{chakraborty_3d_2018}. Moreover, local tunable confinement potential has been realized by Kim and co-workers\cite{wang_electrical_2018} and gate tuning of QD molecules have been shown by Guo and co-workers \cite{zhang_electrotunable_2017}.

There has been significant progress in theoretical understanding of TMDC QDs. Stability and electronic properties of small QDs with various composition, orientation and edge type have been studied within DFT theory \cite{pei_structural_2015, javaid_study_2017, lauritsen_chemistry_2003, lauritsen_size-dependent_2007, mcbride_dft_2009, li_electronic_2007}. Galli and co-workers \cite{li_electronic_2007} studied the electronic properties of  triangular MoS$_\textrm{2}$ quantum dots as a function of the number of layers. 

The ab-initio approaches have also been extended to tight binding models capable of describing quantum dots with lateral sizes up to tens of nanometers. Using a 3-band tight-binding model limited to metal orbitals Peeters and co-workers  \cite{pavlovic_electronic_2015, chen_magnetic_2018} analyzed the effect of quantum dot shape and external magnetic field on the single particle energy spectrum. Using an atomistic tight binding approach spin-valley qubits have been described by Bednarek and co-workers \cite{pawlowski_valley_2018}, Szafran and co-workers \cite{Zebrowski_Szafran_2013, Szafran_Kolasinska_2018, Szafran_Zebrowski_2018} and Guinea and co-workers \cite{chirolli_strain-induced_2019}. In order to understand the size depenence of the electronic states in quantum dots  for realistic sizes involving millions of atoms, $k\cdot p$ and effective  massive Dirac fermion models were applied \cite{kormanyos_spin-orbit_2014,liu_intervalley_2014, brooks_spin-degenerate_2017, szechenyi_impurity-assisted_2018, qu_tunable_2017, dias_robust_2016}. 

In order to realize a spin-valley qubit, a way to control spin and valley properties of electrons in these QDs is needed. Up until now, several means of manipulating the valley index in quantum dots have been studied, such as strain \cite{chirolli_strain-induced_2019}, magnetic field \cite{kormanyos_spin-orbit_2014, brooks_spin-degenerate_2017, szechenyi_impurity-assisted_2018} and coupling to impurity \cite{szechenyi_impurity-assisted_2018}. Valley mixing by the confining potential has also been analyzed by Yao and co-workers\cite{liu_intervalley_2014}. Magnetic control of the spin-valley coupled states in TMDC QDs has been shown by Qu and co-workers \cite{qu_tunable_2017,dias_robust_2016}. Lateral QD molecules have also been studied by several groups \cite{qu_tunable_2011, david_effective_2018}. 

In this work, the states of electrons in quantum dots with millions of atoms are described by the ab-initio based tight-binding Hamiltonian including 3 d-orbitals of metal atoms and 3 p-orbitals of sulfur dimers, made even with respect to the plane of the quantum dot \cite{bieniek_band_2018}. The effect of metallic electrodes is simulated by the parabolic external potential with finite depth and radius, embedded in a computational box up to one million atoms. To avoid edge states associated with a particular termination of the computational box, periodic boundary conditions are used. This allows a study of electrically confined circular quantum dots in TMDCs of experimentally realizable sizes up to 100 nm in radius \cite{pisoni_gate-tunable_2018, wang_electrical_2018}. We find the ladder of degenerate harmonic oscillator states derived from K-valleys, and, as expected and noticed already by Chirolli et al. \cite{chirolli_strain-induced_2019}, two three-fold degenerate harmonic oscillator shells originating from Q points. We also find the splitting of excited harmonic oscillator shells due to the topological moments, opposite for the two valleys \cite{wu_exciton_2015, zhou_berry_2015, srivastava_signatures_2015}. We find the splitting to increase for higher angular momentum shells and to be an order of magnitude higher in Q-derived shells. We also discuss the shell ordering due to  spin orbit coupling (SOC) as well as due to interplay of inter-shell and SOC splitting. These topological and spin splittings together with shell spacing result in the interplay between the K- and Q- derived states which could allow for exploration of the exotic physics of SU(3) symmetry in condensed matter systems \cite{bao_flavor_2019}.

The paper is organized as follows: in Section 2 we describe the tight binding model and the conduction band states of MoS$_\textrm{2}$. In Section 3 we describe the confining potential and the model of MoS$_\textrm{2}$ QD. In Section 4 we present results on the K-derived and Q-derived energy spectrum, shell and spin orbit splitting as well as size-dependent ordering of states in MoS$_\textrm{2}$ QDs. We end with conclusions in Section 5.

\section{The tight binding model and conduction band of MoS$_\textrm{2}$}
We describe here our tight binding model and electronic properties of a single layer of MoS$_\textrm{2}$ \cite{bieniek_band_2018}. We construct the electron's wavefunction as a linear combination of Mo (shown in blue in Fig. \ref{pic2}a) d-orbitals  $m_d=+2$, $m_d=0$, $m_d=-2$ and a linear combination of sulfur dimer (shown in yellow) S$_\textrm{2}$ $p-$ orbitals, even with respect to the plane of Mo atoms, as described in Ref. \onlinecite{bieniek_band_2018}. The nearest and next-nearest neighbour tight-binding Hamiltonian for each spin component can be written as:
\begin{equation}
H_0=\sum_{i\alpha}{\varepsilon_{i\alpha}c_{i\alpha}^+}c_{i\alpha}+\sum_{i\alpha,j\beta}{t_{i\alpha,j\beta}c_{i\alpha}^+c_{j\beta}}, 
\label{ham0}
\end{equation}
where $c_{i\alpha}^+$ describes creation of electron on atom $i$ and orbital $\alpha$ and $t_{i\alpha,j\beta}$ are tunneling matrix elements between atoms $i$ and $j$ and orbitals $\alpha$ and $\beta$,  determined by the Slater-Koster rules. For the metal atom sublattice A and sulfur dimer sublattice B we construct matrix elements $t_{i\alpha,j\beta}$  of the Hamiltonian for nearest neighbour tunneling  $\left\langle {\Psi}_{A,m_d}^{\vec{k}}|H|{\Psi}_{B,m_p}^{\vec{k}}\right\rangle$ and next nearest neighbour tunneling processes, 
$\left\langle{\Psi}_{A,m_d}^{\vec{k}}|H|{\Psi}_{A,m_d}^{\vec{k}}\right\rangle$ and $\left\langle {\Psi}_{B,m_p}^{\vec{k}}|H|{\Psi}_{B,m_p}^{\vec{k}}\right\rangle$, forming a $6\times6$ matrix in the basis of Mo and S$_\textrm{2}$ Bloch functions $\Psi_{A,m_d}^{\vec{k}}=e^{ikr}u_k^{A,m_d}\left(r\right)$ and ${\Psi}_{B,m_p}^{\vec{k}}=e^{ikr}u_k^{B,m_p}\left(r\right)$ of the form:
\scriptsize
\begin{gather*}
H\left(\vec{k} \right)=
\begin{bmatrix}
H_{Mo-Mo} & H_{Mo-S_2} \\
H_{Mo-S_2}^{\dagger} & H_{S_{2}-S_{2}} 
   \end{bmatrix}
   \otimes
\begin{bmatrix}
\mathbb{1} &0 \\
0 & \mathbb{1}
\end{bmatrix}
+
\begin{bmatrix}
H_{SO}(\sigma=1) &0 \\
0 & H_{SO}(\sigma=-1)
\end{bmatrix}
\end{gather*}
\scriptsize
\begin{gather*}
H_{Mo-Mo}=
\begin{bmatrix}
{E_{m_{_{d}}=-2} {+}W_{1}g_{0}(\vec{k})}& W_{3}g_{2}(\vec{k}) & W_{4}g_{4}(\vec{k}) \\
& {E_{m_{_{d}}=0}{+}W_{2}g_{0}(\vec{k})} & W_{3}g_{2}(\vec{k})\\ 
&  & {E_{m_{_{d}}=2}{+}W_{1}g_{0}(\vec{k})} 
   \end{bmatrix}
\end{gather*}
\scriptsize
\begin{gather*}
H_{S_{2}-S_{2}}=
\begin{bmatrix}
{E_{m_{_{p}}=-1}{+}W_{5}g_{0}(\vec{k})} & 0 & W_{7}g_{2}(\vec{k}) \\
& {E_{m_{_{p}}=0}{+}W_{6}g_{0}(\vec{k})} & 0 \\
& & {E_{m_{_{p}}=1}{+}W_{5}g_{0}(\vec{k})} \\
   \end{bmatrix}
\end{gather*}
\scriptsize
\begin{gather*}
H_{Mo-S_{2}}=
\begin{bmatrix}
V_{1}f_{-1}(\vec{k}) & -V_{2}f_{0}(\vec{k}) & V_{3}f_{1}(\vec{k}) \\
-V_{4}f_{0}(\vec{k}) & -V_{5}f_{1}(\vec{k}) & -V_{4}f_{-1}(\vec{k}) \\
-V_{3}f_{1}(\vec{k}) & -V_{2}f_{-1}(\vec{k}) & V_{1}f_{0}(\vec{k}) \\
   \end{bmatrix}\label{6x6}
\end{gather*}
\scriptsize
\begin{gather}
H_{SO}(\sigma)=
\begin{bmatrix}
-\sigma\cdot\lambda_{Mo} & 0 & 0 & 0 & 0 & 0\\
  & 0 & 0 & 0 & 0 & 0 \\
 &  & \sigma\cdot\lambda_{Mo} & 0 & 0 & 0\\
 & & & -\sigma\cdot\frac{\lambda_{S_2}}{2} & 0 & 0 \\
 & & & & 0 & 0 \\
 & & & & & \sigma\cdot\frac{\lambda_{S_2}}{2}
   \end{bmatrix},\label{HSO}
\end{gather}
\normalsize
where the amplitudes $V$, $W$ and k-dependent functions $f$, $g$ are given in the Appendix A of Ref. \onlinecite{bieniek_band_2018} and $\lambda_{Mo}=0.067$ eV, $\lambda_{S_2}=0.02$ eV and $\sigma$ denotes spin index. The diagonal $3\times 3$ blocks correspond to next nearest neighbour Mo-Mo and S$_\textrm{2}$-S$_\textrm{2}$ tunneling while the off-diagonal block translates into Mo-S$_\textrm{2}$ nearest-neighbour tunneling processes.
\begin{figure}[h]
    \centering
    \includegraphics[width=0.45\textwidth]{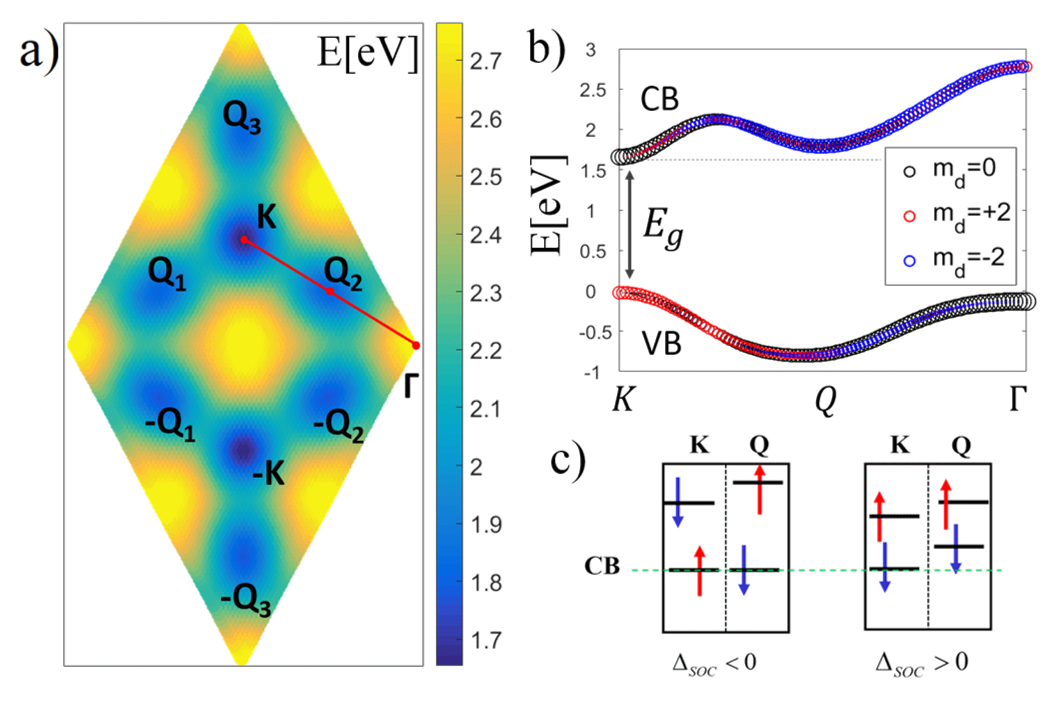}
    \caption{(a) CB energy E$_\textrm{c}$(k) in the BZ consists of K and -K valleys and 6 secondary minima at Q points (note that figure is centered around the K--(-K) edge of the hexagonal Brillouin zone). The red line shows the path along which the CB and VB edges are shown in (b). (c) Two schemes showing possible spin ordering in the CB at K and Q for different TMDCs, e.g. MoS$_\textrm{2}$ (WS$_\textrm{2}$) on the right (left).} \label{pic1}
\end{figure}


We diagonalize the Hamiltonian in Eq. \eqref{HSO} for each k-point in the basis of Bloch functions $\Psi_{A,m_d}^{\vec{k}}$ and ${\Psi}_{B,m_p}^{\vec{k}}$ and obtain 3 conduction and 3 valence band states. The parameters of the Hamiltonian, Eq. \eqref{HSO}, are obtained from the fitting of energy levels to results of ab-initio derived energy bands. \cite{kadantsev_electronic_2012,bieniek_band_2018}

The lowest energy conduction band (CB) dispersion E$_\textrm{c}$(k) in the first Brillouin zone (BZ) is shown in Fig. \ref{pic1}a. The Brillouin zone is hexagonal, with 6 K points at the six corners, with 3 of them being equivalent up to a reciprocal lattice vector translation in both K and -K valleys. The lack of inversion symmetry in the unit cell leads to K and -K points being non-equivalent.  A single layer of TMDCs has a direct band gap, located in the K and -K points. Three secondary Q minima exist around K and -K valleys.  Fig. \ref{pic1}b shows the valence and conduction bands for MoS$_\textrm{2}$ plotted along the red path shown in Fig. \ref{pic1}a. The additional Q conduction minima along K-$\Gamma$ line are responsible for nesting of the conduction and valence bands. The low energy bands are mainly composed of Mo d-orbitals, with $m_d=0$ building the bottom of the conduction band at K and $m_d=+2$ contributing to the top of the valence band at K. The $m_d=0$ orbital contributes to the conduction band at K, while at Q point a different orbital, $m_d=-2$, contributes to the conduction band. Hence, the QD states obtained below will derive from the conduction band states, with both K and Q minima, with their corresponding d-orbitals, contributing to these states.

The spin-orbit coupling plays an important role in TMDCs, resulting in spin splitting reaching up to 130-145 meV in the valence band and 3-4 meV in conduction band for MoS$_\textrm{2}$ \cite{kadantsev_electronic_2012, zhang_direct_2014, marinov_resolving_2017}. Due to spin-orbit coupling, conduction band edges in some TMDCs can be built from states in the vicinity of both K and Q points, when the spin up states at K and spin down states at Q become degenerate (Fig. \ref{pic1}c left). This scheme prevails in materials with tungsten as a metal \cite{kosmider_large_2013}. For compounds with molybdenum (Fig. \ref{pic1}c right) a gap between K - and Q - points spin-split bands is larger \cite{kormanyos_monolayer_2013, kosmider_large_2013, yu_nonlinear_2014}. In this work we focus on MoS$_\textrm{2}$, but we explore the physics of QD states built from K and Q points, which may be equally relevant for the low energy spectra in different MX$_\textrm{2}$ materials.

\section{The model of a quantum dot}
\begin{figure}[h]
    \centering
    \includegraphics[width=0.45\textwidth]{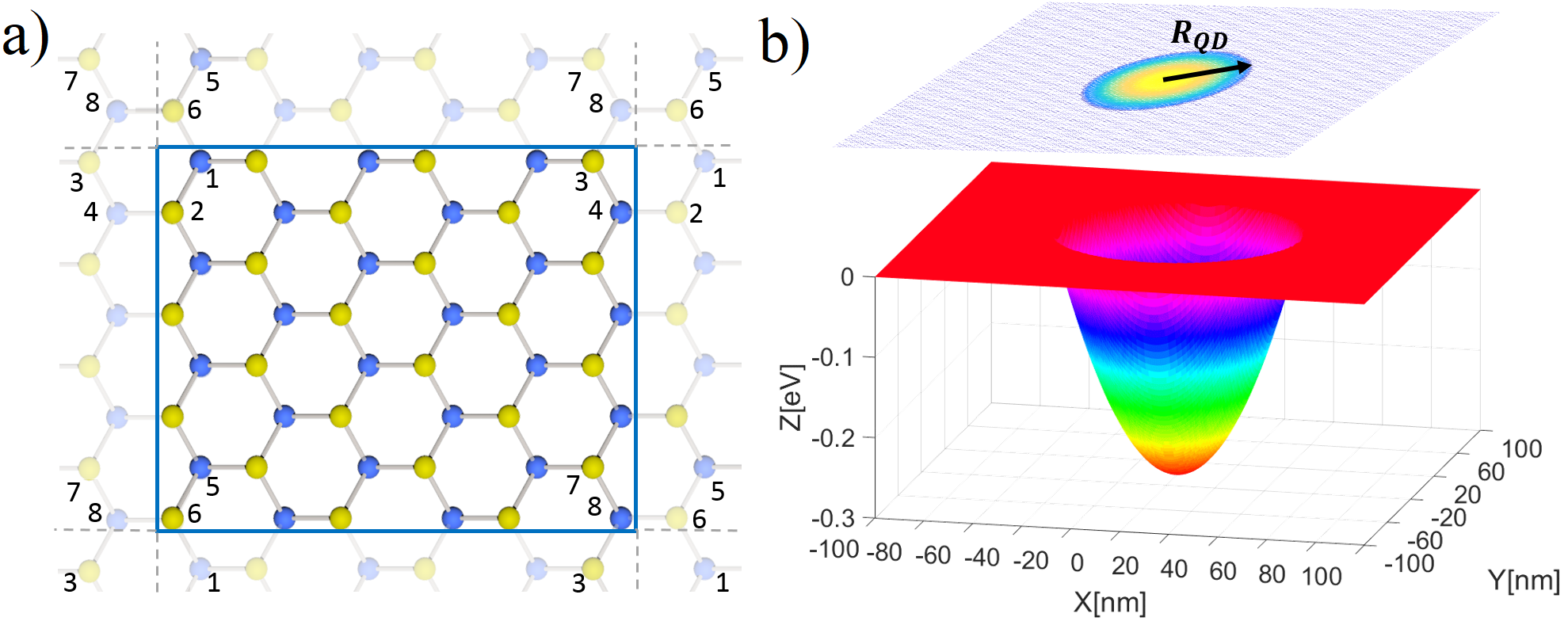}
    \caption{(a) Rectangular computation box of the MoS$_\textrm{2}$ plane with periodic boundary conditions. (b) Profile of the parabolic confining potential forming a QD of radius $R_{QD}$.} \label{pic2}
\end{figure}

We now discuss our model of a quantum dot. We start with a rectangular computational box of a single plane of MoS$_\textrm{2}$ with periodic boundary conditions as described in Section 2 and shown in Fig. \ref{pic2}a.  We then introduce a parabolic potential generated by metallic gates\cite{kyriakidis_prb2002} as shown in Fig. \ref{pic2}b. 
The  metallic gates introduce an electric field perpendicular to the atomic layers as studied in Refs. \onlinecite{klein_stark_2016, chu_electrically_2015}. For typical applied voltages and splittings off the even and odd sulfur orbitals at the K-point we estimate the admixture of odd orbitals into even orbitals induced by the metallic gates to be under 1{\%}. Hence the total Hamiltonian of the parabolic QD (Fig. \ref{pic2}a) with radius $R_{QD}$ is given by the Hamiltonian $H_0$ describing even orbitals and the external potential $V$:
\begin{equation}
 H=H_0+\sum_{i\alpha} V_ic_{i\alpha}^+c_{i\alpha}, \label{ham}
\end{equation}
where $V_i$ is the external potential on atom $i$ generated by metallic gates. For gated  quantum dots the potential $V(r)$ is largely parabolic and given by \cite{kyriakidis_prb2002}:
\begin{equation}
V_i=V(r_i)=\left\{
\begin{aligned}
\frac{1}{2}\omega^2 r_i^2-V_{max}, & \text{ for } r_i<R_{QD}\\
0\quad \quad \quad \quad \quad, & \text{ for } r_i>R_{QD} .
\end{aligned}
\right. \label{dot}
\end{equation}
The parabolic confining potential can be expressed by the corresponding harmonic oscillator level spacing $\omega=2|V_{max}|/R_{QD}^2$  defined by an electrostatic potential with depth $V_{max}$ and radius $R_{QD}$. For definiteness, we keep $V_{max}$ at 300 meV throughout this work. At the boundary of the dot, the confining potential goes to 0. Dot edges are kept sufficiently far from the computational box edges, connected by periodic boundary conditions (BC). We have confirmed that in our model states localized inside the dot are not influenced by the choice of BC. The sizes of the computational domain  studied are up to $\sim220\times220$ nm, which corresponds to $\sim1.1\cdot10^6$ atoms, and up to a 100 nm dot radius, corresponding to experimentally studied systems \cite{zhang_electrotunable_2017, pisoni_gate-tunable_2018}.

Diagonalization of such large, sparse Hamiltonian matrices, is performed using the FEAST algorithm \cite{PhysRevB.79.115112} as well as with sparse matrix diagonalization routines within the PETSC library \cite{petsc-web-page}.

\section{Results}    

\subsection{K-point-derived and Q-point-derived spectrum of electronic states}

\begin{figure}[ht]
    \centering
    \includegraphics[width=0.45\textwidth]{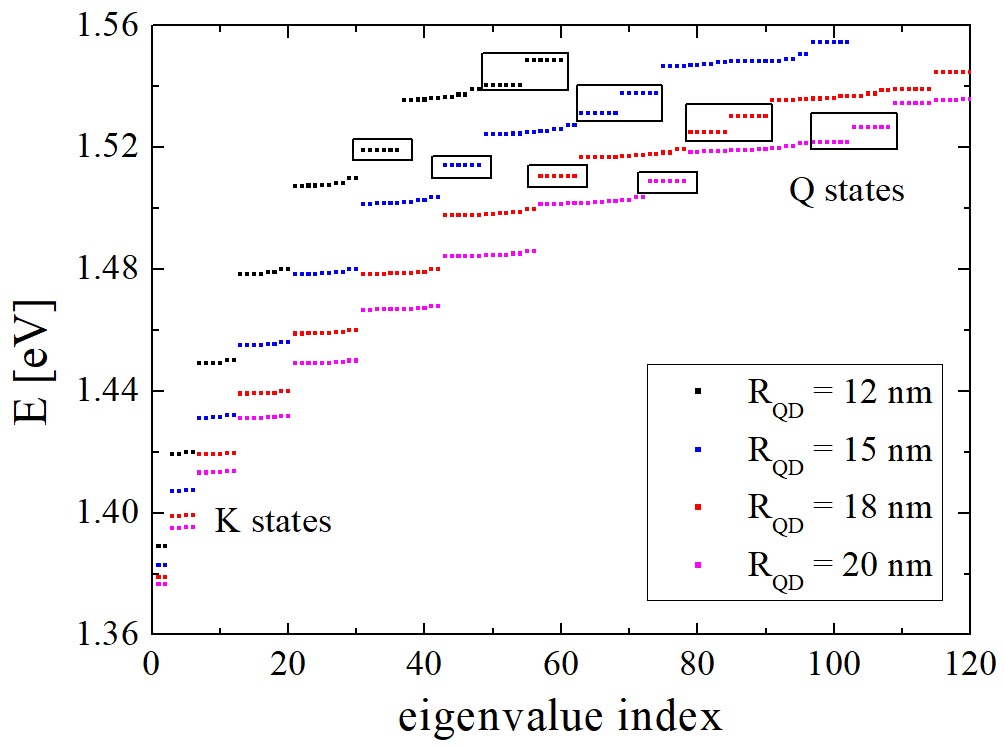}
    \caption{QD states (with no SOC) for increasing QD radius $R_{QD}$. K-derived harmonic oscillator ladder of states is interrupted by the Q-derived states marked with boxes higher in the spectrum.} \label{pic4}
\end{figure}

\begin{figure}[h]
    \centering
    \includegraphics[width=0.45\textwidth]{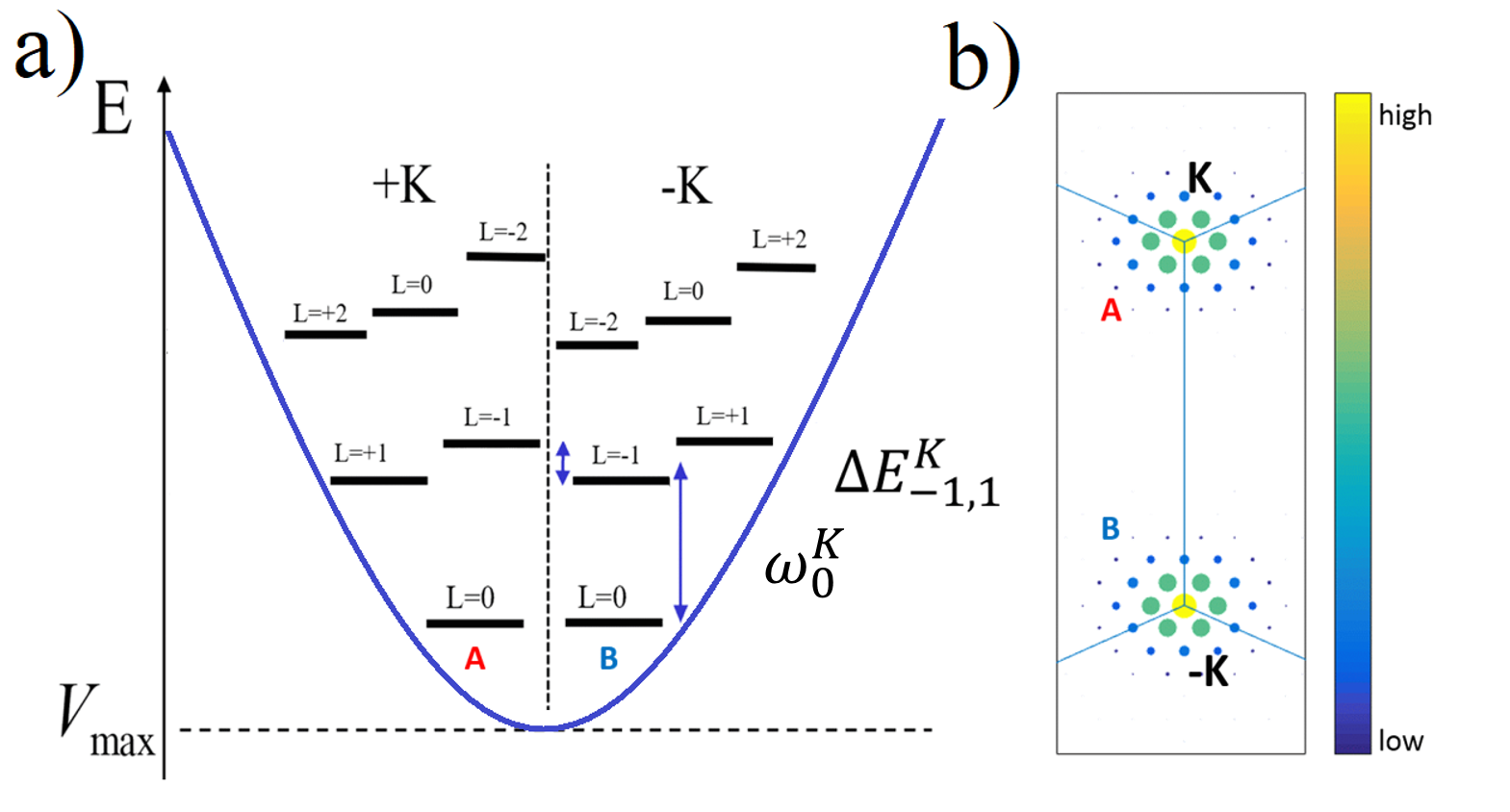}
    \caption{(a) Ladder of QD K-derived states (parabola depicts the confining potential). Harmonic oscillator shells are doubled due to valley and due to spin (spin degeneracy not shown). Inter-shell spacing is labeled with $\omega_0^K$, while the intra-shell splitting is labeled with $\Delta E_{-1,1}^K$. Angular momentum of states is denoted by L. (b) Fourier composition of the two states from the lowest shell marked with A \& B.} \label{figK}
\end{figure}

We start with a parabolic QD defined electrostatically on representative TMDC, MoS$_\textrm{2}$ , as shown in Fig. \ref{pic2}b. 
For  clarity, we first neglect spin-orbit coupling (SOC) in the tight-binding Hamiltonian in Eq. \eqref{HSO}.  The results of diagonalization of the quantum dot Hamiltonian with $V_{max}=300$ meV and variable $R_{QD}=\{12,15,18,20\}$ nm are shown in Fig. \ref{pic4}. We see that electronic states are arranged into almost equally spaced electronic shells. Each shell consists of states derived from K and -K points, doubly degenerate due to spin, as schematically shown in Fig. \ref{figK}a. Fig. \ref{figK}b shows the Fourier composition of the first 2-level shell of the QD. With very small spin-orbit splitting one can attribute each of these 2 states to either +K or -K valley. In each valley there are equally spaced electronic shells with degeneracies identical to the spectrum of two harmonic oscillators as observed directly in self-assembled quantum dots \cite{raymond_excitonic_2004}. However, unlike in GaAs or self-assembled quantum dots, the degeneracy of each electronic shell is removed, an effect discussed below.    

Fig. \ref{pic4} shows the evolution of the energy levels with increasing dot radius $R_{QD}$ while keeping the depth of potential fixed. We see that with increasing $R_{QD}$ more electronic shells are confined within the dot. However, in contrast with gated quantum dots in GaAs, for all studied QD sizes in addition to K derived electronic shells, there exists perfectly 6-fold degenerate shells, emerging at higher energy and marked with rectangular boxes in Fig. \ref{pic4}. 

\begin{figure}[ht]
    \centering
    \includegraphics[width=0.45\textwidth]{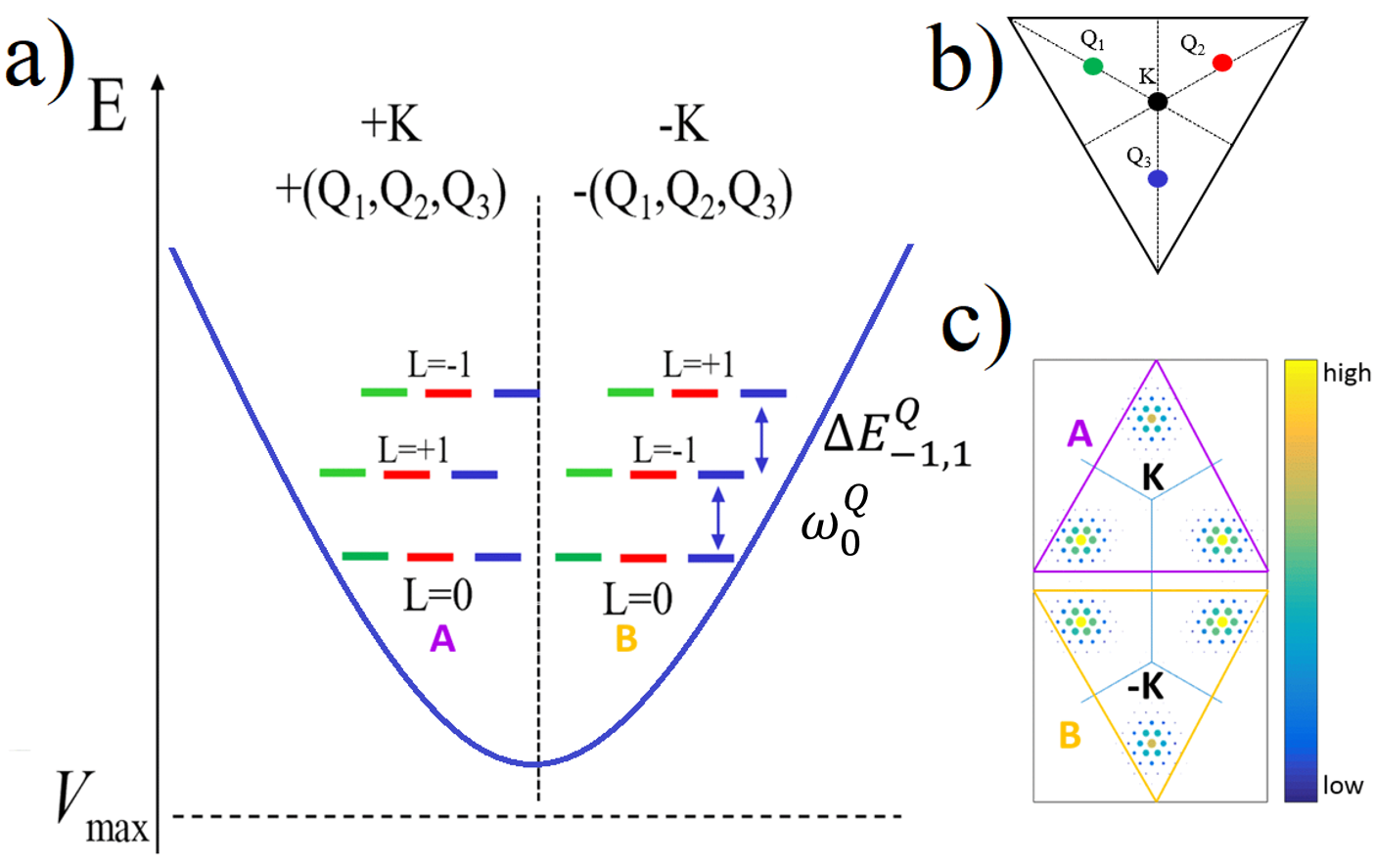}
    \caption{(a) Ladder of QD Q-derived states (parabola depicts the confining potential). Harmonic oscillator shells are six-fold degenerate due to K and -K valley and 3 Q points around each valley as shown in matching colours in (b) (spin degeneracy not shown). Inter-shell spacing is labeled with $\omega_0^Q$, while the intra-shell splitting is labeled with $\Delta E_{-1,1}^Q$. (c) Fourier composition of the two sets of states from the lowest shell is marked with A \& B.} \label{figQ}
\end{figure}

The 6-fold degeneracy of new electronic shells stems from the 3 non-equivalent Q points around the K valley and the 3 non-equivalent Q points around the  -K valley, as shown schematically in Fig. \ref{figQ}a and \ref{figQ}b. Fig \ref{figQ}c shows the Fourier composition of the first shell of 6 degenerate Q-derived states. For very small spin orbit coupling two sets of 3 states can be attributed to the mixture of 3 Q points around the K and -K valley. Interestingly, these Q-derived shells can be understood as the condensed matter physics analogue of flavour SU(3) symmetry \cite{bao_flavor_2019}, describing quarks in high energy physics.

\subsection{Topological splitting of electronic shells}
In spectra shown in Fig. \ref{pic4} we observe intra-shell splitting despite cylindrical symmetry of the confining potential. The splitting appears to depend on the angular momentum of harmonic oscillator states in the degenerate electronic shell. As shown experimentally in Ref. \cite{raymond_excitonic_2004} the application of an external magnetic field removes the degeneracy of harmonic oscillator states. Hence, this splitting can be understood as resulting from Berry's curvature, analogous to a magnetic field acting on the finite angular momentum states,  in opposite directions in K and -K valleys \cite{srivastava_signatures_2015, zhou_berry_2015}. 

As shown schematically in Fig. \ref{figK}a and Fig. \ref{figQ}a, this splitting is observed for both K- and Q-derived harmonic oscillator shells. We note that for the same $R_{QD}$=30 nm, intra-shell "topological" splitting grows with shell number and, importantly, is an order of magnitude stronger for Q-point states, reaching up to 6.5 meV. We note that the smaller the dot, the larger the splitting is observed. We notice also that angular momentum $L=\pm2$ state splitting around $L=0$ state for K-point series, is not symmetric, suggesting that Berry's curvature might influence also the L=0 states, as in the s-series of excitons in TMDC materials \cite{trushin_model_2018, bieniek_to_be_published}.

\begin{figure}[ht]
    \centering
    \includegraphics[width=0.35\textwidth]{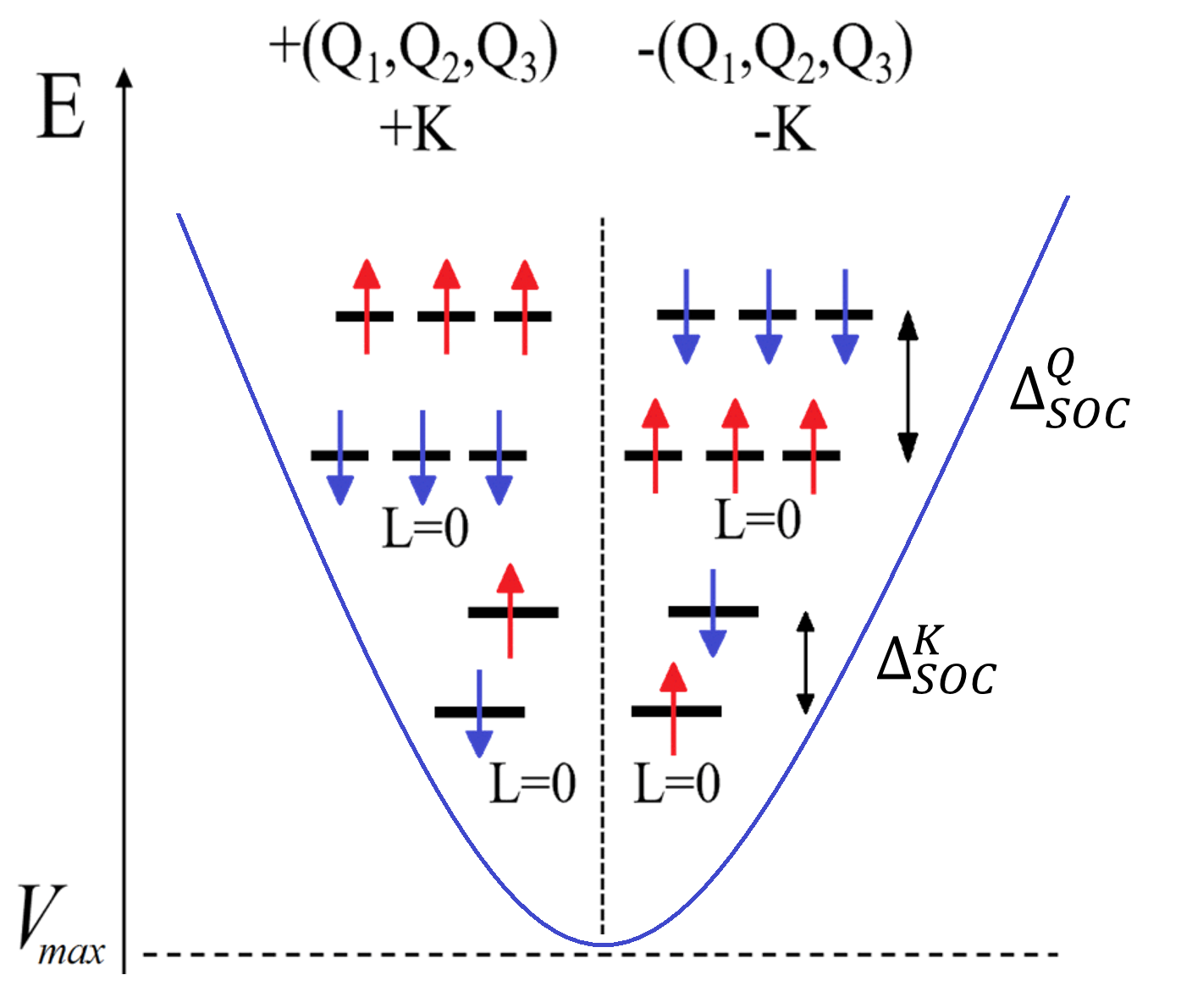}
    \caption{Diagram of the spin ordering of levels from K-derived and Q-derived ladder (parabola depicts the confining potential). Spins order oppositely in levels from around K and -K valley. Spin splittings of K-derived (Q-derived) shells are marked with $\Delta^K_{SOC}$ ($\Delta^Q_{SOC}$).} \label{pic6}
\end{figure}

\subsection{Spin-orbit coupling vs. shell splitting}
We now turn on the spin-orbit coupling (SOC) in the TB Hamiltonian given by Eq. \eqref{HSO}, which induces a splitting between spin up and spin down states in all shells, as shown schematically in Fig \ref{pic6}. The splitting 
$\Delta_{\textrm{SOC}}$ changes sign when going from K to -K valley. It increases with QD radius $R_{QD}$ and for the K-derived states it reaches value close to the bulk value of 4.2 meV \cite{bieniek_band_2018, kadantsev_electronic_2012} for $R_{QD}=100$ nm.

\begin{figure}[h]
    \centering
    \includegraphics[width=0.48\textwidth]{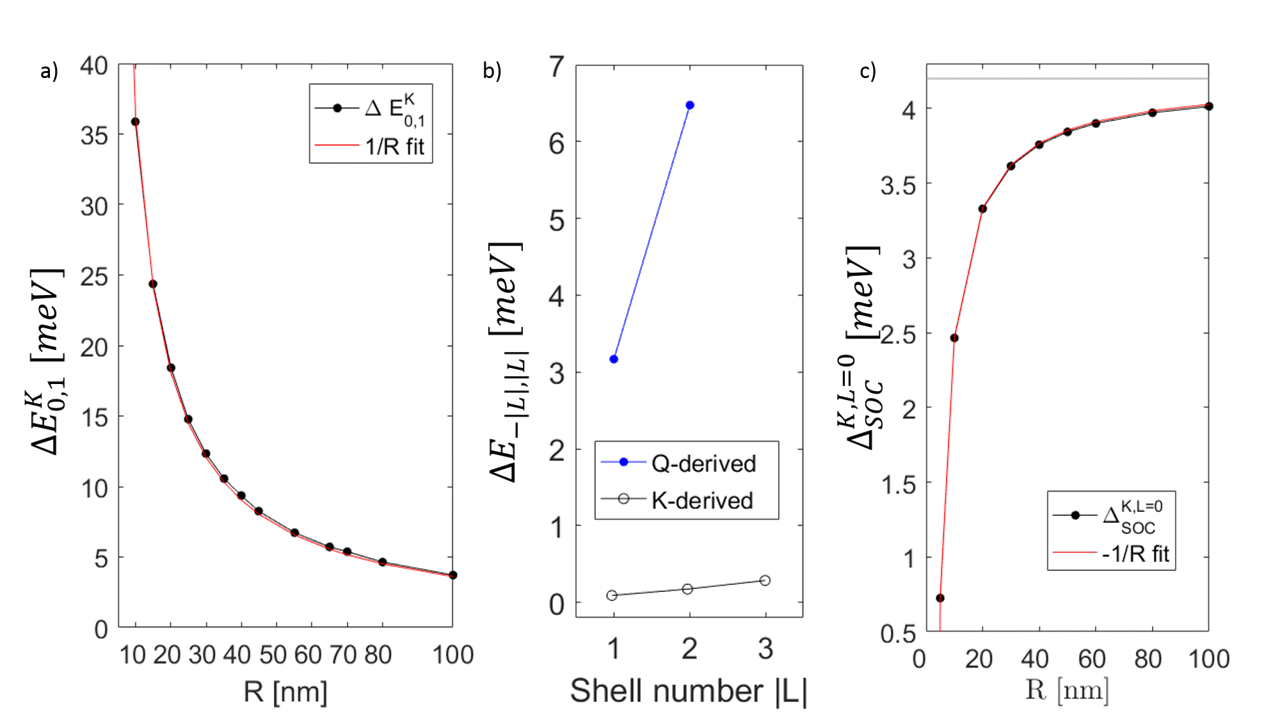}
    \caption{Inter-shell, intra-shell and SO splittings  as a function of quantum dot radius in MoS$_\textrm{2}$ QDs. (a) Inter-shell spacing decreases inversely proportionally to QD radius $R_{QD}$. (b) Intra-shell topological splitting increases for higher shells and for Q-derived shells it reaches 6.5 meV for $R_{QD}=30$ nm. (c) SOC splitting increases for larger dots as -$1/ R_{QD}$ and saturates close to bulk value of $4.2$ meV for QDs larger than $R_{QD}=100$ nm. For $R_{QD}=100$ nm the SOC splitting is higher than the inter-shell spacing, which affects the order of levels.} \label{pic7}
\end{figure}

Fig. \ref{pic7} shows the behaviour of the inter-shell, intra-shell and SO splittings in MoS$_{\textrm{2}}$ QDs. As shown in Fig. \ref{pic7}a, the splitting between the first and second shell of K-derived states decreases  inversely proportionally to the QD radius $R_{QD}$, as expected for a harmonic oscillator and as seen previously for GaAs QD \cite{raymond_excitonic_2004, ciorga_addition_2000}. However, unlike in GaAs QDs, the TMDC QD spectrum is also determined by the topological intra-shell splitting, which grows with angular momentum of the shell for both K- and Q- derived states, as shown in Fig. \ref{pic7}b. Large intra-shell splittings in the Q-derived shells have no counterpart in III-V semiconductor nanostructures.

Importantly, the QD energy spectrum also depends heavily on the SO splitting. As shown in Fig. \ref{pic7}c the SO splitting in the first K-derived shell of states grows with QD size as -$1/ R_{QD}$ and saturates for systems close to bulk size value of 4.2 meV,  marked with a grey line in Fig. \ref{pic7}c. This interplay of splittings will determine the order of shells for TMDC QDs and therefore, the shell filling in a many-electron system.

\begin{figure}[ht]
    \centering
    \includegraphics[width=0.4\textwidth]{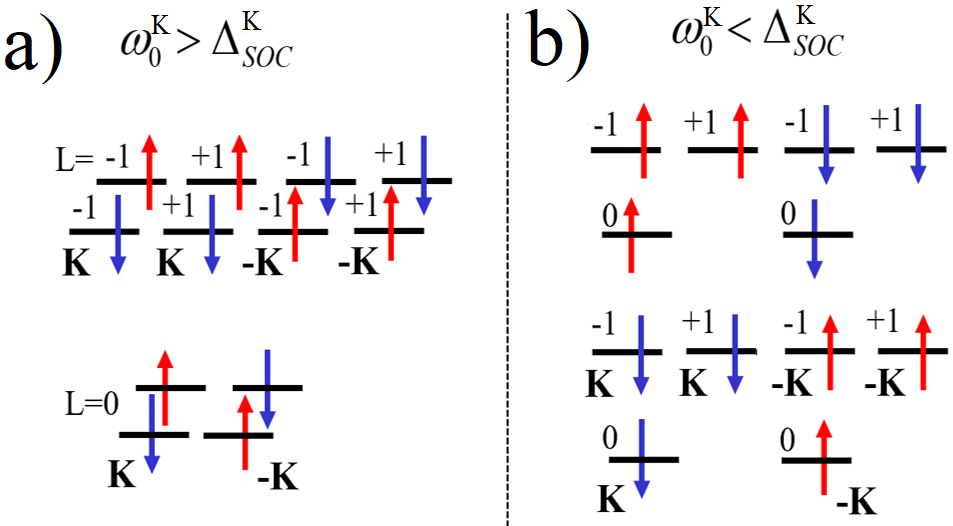}
    \caption{Two regimes for inter-shell spacing $\omega_0^K$ relative to the SOC splitting $\Delta_{SOC}^K$. (a) If $\omega_0^K$ is large compared to $\Delta_{SOC}^K$, the shells stay separate. (b) If $\omega_0^K$ is small compared to $\Delta_{SOC}^K$, the shells intertwine.} \label{pic8}
\end{figure}

In Fig. \ref{pic8} we show two scenarios of the order of K-derived shells for $\Delta_{SOC}^K > \omega_0^K$ and $\Delta_{SOC}^K < \omega_0^K$ type of materials. When $\omega_0^K > \Delta_{SOC}^K$ (Fig. \ref{pic8}a), the lower energy shells are ordered according to angular momentum $L$. First two energies are doubly degenerate, and the  fifth state belongs to the next $L=1$ shell. However, when $\omega_0^K < \Delta_{SOC}^K$ (Fig. \ref{pic8}b), the energy of the third state already reaches the energy of the $L=1$ shell. 

Interestingly, this reordering can be also observed for MoS$_\textrm{2}$ QDs, if the $R_{QD}$ can be varied. As can be seen from Fig. \ref{pic7}a and Fig. \ref{pic7}c, for MoS$_\textrm{2}$ QDs with radii larger than 100 nm the inter-shell splitting of the lowest K-derived shell $\omega_0^K$ is lower than its SO splitting $\Delta_{SOC}^{K}$, which mixes the order of the shell spectrum, like in $\Delta_{SOC}^K < \omega_0^K$ type of TMDCs. By fabricating QDs with two sizes it is possible to realize the scenarios described in Fig. \ref{pic8}, mimicking two distinct TMDC compounds. 

\section{Conclusions}
In this work we presented an atomistic theory of electrons confined by metallic gates in a single layer of transition metal dichalcogenides. The electronic states were described by the tight-binding model including metal and sulfur orbitals and computed using a computational box including up to one million atoms with periodic boundary conditions and with embedded in it a parabolic confining potential due to external gates. This allowed us  to determine the energy spectrum in quantum dots with experimentally relevant sizes. We found a two-fold valley - degenerate energy spectrum and a six-fold degenerate spectrum associated with Q-valleys. We discussed the role of spin splitting and topological moments on the K and Q valley electronic states. We pointed out importance of SU(3) flavor Q – point states for low lying QD states. Future work will determine means of controlling the valley degree of freedom and the role of electron - electron interactions. 

\section*{Acknowledgments}
M.B., L.S., and P.H. thank M. Korkusinski, Y. Saleem, M. Cygorek, A. Luican-Mayer, A. Badolato, I. Ozfidan, L. Gaudreau, S. Studenikin   and A. Sachrajda for discussions. M.B., L.S., and P.H. acknowledge support from NSERC Discovery and QC2DM Strategic Project grants as well as uOttawa Research Chair in Quantum Theory of Materials, Nanostructures and Devices. M.B. acknowledges financial support from National Science Center (NCN), Poland, grant Maestro No. 2014/14/A/ST3/00654. Computing resources from Compute Canada and Wroclaw Center for Networking and Supercomputing are gratefully acknowledged. 

\bibliography{bib1}

\begin{thebibliography}{99}%
\makeatletter
\providecommand \@ifxundefined [1]{%
 \@ifx{#1\undefined}
}%
\providecommand \@ifnum [1]{%
 \ifnum #1\expandafter \@firstoftwo
 \else \expandafter \@secondoftwo
 \fi
}%
\providecommand \@ifx [1]{%
 \ifx #1\expandafter \@firstoftwo
 \else \expandafter \@secondoftwo
 \fi
}%
\providecommand \natexlab [1]{#1}%
\providecommand \enquote  [1]{``#1''}%
\providecommand \bibnamefont  [1]{#1}%
\providecommand \bibfnamefont [1]{#1}%
\providecommand \citenamefont [1]{#1}%
\providecommand \href@noop [0]{\@secondoftwo}%
\providecommand \href [0]{\begingroup \@sanitize@url \@href}%
\providecommand \@href[1]{\@@startlink{#1}\@@href}%
\providecommand \@@href[1]{\endgroup#1\@@endlink}%
\providecommand \@sanitize@url [0]{\catcode `\\12\catcode `\$12\catcode
  `\&12\catcode `\#12\catcode `\^12\catcode `\_12\catcode `\%12\relax}%
\providecommand \@@startlink[1]{}%
\providecommand \@@endlink[0]{}%
\providecommand \url  [0]{\begingroup\@sanitize@url \@url }%
\providecommand \@url [1]{\endgroup\@href {#1}{\urlprefix }}%
\providecommand \urlprefix  [0]{URL }%
\providecommand \Eprint [0]{\href }%
\providecommand \doibase [0]{http://dx.doi.org/}%
\providecommand \selectlanguage [0]{\@gobble}%
\providecommand \bibinfo  [0]{\@secondoftwo}%
\providecommand \bibfield  [0]{\@secondoftwo}%
\providecommand \translation [1]{[#1]}%
\providecommand \BibitemOpen [0]{}%
\providecommand \bibitemStop [0]{}%
\providecommand \bibitemNoStop [0]{.\EOS\space}%
\providecommand \EOS [0]{\spacefactor3000\relax}%
\providecommand \BibitemShut  [1]{\csname bibitem#1\endcsname}%
\let\auto@bib@innerbib\@empty
\bibitem [{\citenamefont {Brum}\ and\ \citenamefont
  {Hawrylak}(1997)}]{brum_coupled_1997}%
  \BibitemOpen
  \bibfield  {author} {\bibinfo {author} {\bibfnamefont {J.~A.}\ \bibnamefont
  {Brum}}\ and\ \bibinfo {author} {\bibfnamefont {P.}~\bibnamefont
  {Hawrylak}},\ }\href {\doibase http://dx.doi.org/10.1006/spmi.1996.0263}
  {\bibfield  {journal} {\bibinfo  {journal} {Superlattices and
  Microstructures}\ }\textbf {\bibinfo {volume} {22}},\ \bibinfo {pages} {431 }
  (\bibinfo {year} {1997})}\BibitemShut {NoStop}%
\bibitem [{\citenamefont {Loss}\ and\ \citenamefont
  {DiVincenzo}(1998)}]{loss_quantum_1998}%
  \BibitemOpen
  \bibfield  {author} {\bibinfo {author} {\bibfnamefont {D.}~\bibnamefont
  {Loss}}\ and\ \bibinfo {author} {\bibfnamefont {D.~P.}\ \bibnamefont
  {DiVincenzo}},\ }\href {\doibase 10.1103/PhysRevA.57.120} {\bibfield
  {journal} {\bibinfo  {journal} {Physical Review A}\ }\textbf {\bibinfo
  {volume} {57}},\ \bibinfo {pages} {120} (\bibinfo {year} {1998})}\BibitemShut
  {NoStop}%
\bibitem [{\citenamefont {Tarucha}\ \emph {et~al.}(1996)\citenamefont
  {Tarucha}, \citenamefont {Austing}, \citenamefont {Honda}, \citenamefont
  {van~der Hage},\ and\ \citenamefont
  {Kouwenhoven}}]{tarucha_shellfilling_1996}%
  \BibitemOpen
  \bibfield  {author} {\bibinfo {author} {\bibfnamefont {S.}~\bibnamefont
  {Tarucha}}, \bibinfo {author} {\bibfnamefont {D.~G.}\ \bibnamefont
  {Austing}}, \bibinfo {author} {\bibfnamefont {T.}~\bibnamefont {Honda}},
  \bibinfo {author} {\bibfnamefont {R.~J.}\ \bibnamefont {van~der Hage}}, \
  and\ \bibinfo {author} {\bibfnamefont {P.}~\bibnamefont {Kouwenhoven},
  \bibfnamefont {L.}},\ }\href@noop {} {\bibfield  {journal} {\bibinfo
  {journal} {Physical Review Letters}\ }\textbf {\bibinfo {volume} {77}},\
  \bibinfo {pages} {3613} (\bibinfo {year} {1996})}\BibitemShut {NoStop}%
\bibitem [{\citenamefont {Ciorga}\ \emph {et~al.}(2000)\citenamefont {Ciorga},
  \citenamefont {Sachrajda}, \citenamefont {Hawrylak}, \citenamefont {Gould},
  \citenamefont {Zawadzki}, \citenamefont {Jullian}, \citenamefont {Feng},\
  and\ \citenamefont {Wasilewski}}]{ciorga_addition_2000}%
  \BibitemOpen
  \bibfield  {author} {\bibinfo {author} {\bibfnamefont {M.}~\bibnamefont
  {Ciorga}}, \bibinfo {author} {\bibfnamefont {A.~S.}\ \bibnamefont
  {Sachrajda}}, \bibinfo {author} {\bibfnamefont {P.}~\bibnamefont {Hawrylak}},
  \bibinfo {author} {\bibfnamefont {C.}~\bibnamefont {Gould}}, \bibinfo
  {author} {\bibfnamefont {P.}~\bibnamefont {Zawadzki}}, \bibinfo {author}
  {\bibfnamefont {S.}~\bibnamefont {Jullian}}, \bibinfo {author} {\bibfnamefont
  {Y.}~\bibnamefont {Feng}}, \ and\ \bibinfo {author} {\bibfnamefont
  {Z.}~\bibnamefont {Wasilewski}},\ }\href {\doibase
  10.1103/PhysRevB.61.R16315} {\bibfield  {journal} {\bibinfo  {journal}
  {Physical Review B}\ }\textbf {\bibinfo {volume} {61}},\ \bibinfo {pages}
  {R16315} (\bibinfo {year} {2000})}\BibitemShut {NoStop}%
\bibitem [{\citenamefont {Meier}\ \emph {et~al.}(2003)\citenamefont {Meier},
  \citenamefont {Levy},\ and\ \citenamefont {Loss}}]{meier_quantum_2003}%
  \BibitemOpen
  \bibfield  {author} {\bibinfo {author} {\bibfnamefont {F.}~\bibnamefont
  {Meier}}, \bibinfo {author} {\bibfnamefont {J.}~\bibnamefont {Levy}}, \ and\
  \bibinfo {author} {\bibfnamefont {D.}~\bibnamefont {Loss}},\ }\href {\doibase
  10.1103/PhysRevLett.90.047901} {\bibfield  {journal} {\bibinfo  {journal}
  {Physical Review Letters}\ }\textbf {\bibinfo {volume} {90}},\ \bibinfo
  {pages} {047901} (\bibinfo {year} {2003})}\BibitemShut {NoStop}%
\bibitem [{\citenamefont {Elzerman}\ \emph {et~al.}(2004)\citenamefont
  {Elzerman}, \citenamefont {Hanson}, \citenamefont {Willems~van Beveren},
  \citenamefont {Witkamp}, \citenamefont {Vandersypen},\ and\ \citenamefont
  {Kouwenhoven}}]{elzerman_single-shot_2004}%
  \BibitemOpen
  \bibfield  {author} {\bibinfo {author} {\bibfnamefont {J.~M.}\ \bibnamefont
  {Elzerman}}, \bibinfo {author} {\bibfnamefont {R.}~\bibnamefont {Hanson}},
  \bibinfo {author} {\bibfnamefont {L.~H.}\ \bibnamefont {Willems~van
  Beveren}}, \bibinfo {author} {\bibfnamefont {B.}~\bibnamefont {Witkamp}},
  \bibinfo {author} {\bibfnamefont {L.~M.~K.}\ \bibnamefont {Vandersypen}}, \
  and\ \bibinfo {author} {\bibfnamefont {L.~P.}\ \bibnamefont {Kouwenhoven}},\
  }\href {\doibase 10.1038/nature02693} {\bibfield  {journal} {\bibinfo
  {journal} {Nature}\ }\textbf {\bibinfo {volume} {430}},\ \bibinfo {pages}
  {431} (\bibinfo {year} {2004})}\BibitemShut {NoStop}%
\bibitem [{\citenamefont {Hanson}\ \emph {et~al.}(2007)\citenamefont {Hanson},
  \citenamefont {Kouwenhoven}, \citenamefont {Petta}, \citenamefont {Tarucha},\
  and\ \citenamefont {Vandersypen}}]{hanson_spins_2007}%
  \BibitemOpen
  \bibfield  {author} {\bibinfo {author} {\bibfnamefont {R.}~\bibnamefont
  {Hanson}}, \bibinfo {author} {\bibfnamefont {L.~P.}\ \bibnamefont
  {Kouwenhoven}}, \bibinfo {author} {\bibfnamefont {J.~R.}\ \bibnamefont
  {Petta}}, \bibinfo {author} {\bibfnamefont {S.}~\bibnamefont {Tarucha}}, \
  and\ \bibinfo {author} {\bibfnamefont {L.~M.~K.}\ \bibnamefont
  {Vandersypen}},\ }\href {\doibase 10.1103/RevModPhys.79.1217} {\bibfield
  {journal} {\bibinfo  {journal} {Reviews of Modern Physics}\ }\textbf
  {\bibinfo {volume} {79}},\ \bibinfo {pages} {1217} (\bibinfo {year}
  {2007})}\BibitemShut {NoStop}%
\bibitem [{\citenamefont {Kyriakidis}\ \emph {et~al.}(2002)\citenamefont
  {Kyriakidis}, \citenamefont {Pioro-Ladriere}, \citenamefont {Ciorga},
  \citenamefont {Sachrajda},\ and\ \citenamefont
  {Hawrylak}}]{kyriakidis_prb2002}%
  \BibitemOpen
  \bibfield  {author} {\bibinfo {author} {\bibfnamefont {J.}~\bibnamefont
  {Kyriakidis}}, \bibinfo {author} {\bibfnamefont {M.}~\bibnamefont
  {Pioro-Ladriere}}, \bibinfo {author} {\bibfnamefont {M.}~\bibnamefont
  {Ciorga}}, \bibinfo {author} {\bibfnamefont {A.~S.}\ \bibnamefont
  {Sachrajda}}, \ and\ \bibinfo {author} {\bibfnamefont {P.}~\bibnamefont
  {Hawrylak}},\ }\href {\doibase 10.1103/PhysRevB.66.035320} {\bibfield
  {journal} {\bibinfo  {journal} {Phys. Rev. B}\ }\textbf {\bibinfo {volume}
  {66}},\ \bibinfo {pages} {035320} (\bibinfo {year} {2002})}\BibitemShut
  {NoStop}%
\bibitem [{\citenamefont {Hsieh}\ \emph {et~al.}(2012)\citenamefont {Hsieh},
  \citenamefont {Shim}, \citenamefont {Korkusinski},\ and\ \citenamefont
  {Hawrylak}}]{hsieh_physics_2012}%
  \BibitemOpen
  \bibfield  {author} {\bibinfo {author} {\bibfnamefont {C.-Y.}\ \bibnamefont
  {Hsieh}}, \bibinfo {author} {\bibfnamefont {Y.-P.}\ \bibnamefont {Shim}},
  \bibinfo {author} {\bibfnamefont {M.}~\bibnamefont {Korkusinski}}, \ and\
  \bibinfo {author} {\bibfnamefont {P.}~\bibnamefont {Hawrylak}},\ }\href
  {\doibase 10.1088/0034-4885/75/11/114501} {\bibfield  {journal} {\bibinfo
  {journal} {Reports on Progress in Physics}\ }\textbf {\bibinfo {volume}
  {75}},\ \bibinfo {pages} {114501} (\bibinfo {year} {2012})}\BibitemShut
  {NoStop}%
\bibitem [{\citenamefont {Nowack}\ \emph {et~al.}(2011)\citenamefont {Nowack},
  \citenamefont {Shafiei}, \citenamefont {Laforest}, \citenamefont
  {Prawiroatmodjo}, \citenamefont {Schreiber}, \citenamefont {Reichl},
  \citenamefont {Wegscheider},\ and\ \citenamefont
  {Vandersypen}}]{nowack_single-shot_2011}%
  \BibitemOpen
  \bibfield  {author} {\bibinfo {author} {\bibfnamefont {K.~C.}\ \bibnamefont
  {Nowack}}, \bibinfo {author} {\bibfnamefont {M.}~\bibnamefont {Shafiei}},
  \bibinfo {author} {\bibfnamefont {M.}~\bibnamefont {Laforest}}, \bibinfo
  {author} {\bibfnamefont {G.~E. D.~K.}\ \bibnamefont {Prawiroatmodjo}},
  \bibinfo {author} {\bibfnamefont {L.~R.}\ \bibnamefont {Schreiber}}, \bibinfo
  {author} {\bibfnamefont {C.}~\bibnamefont {Reichl}}, \bibinfo {author}
  {\bibfnamefont {W.}~\bibnamefont {Wegscheider}}, \ and\ \bibinfo {author}
  {\bibfnamefont {L.~M.~K.}\ \bibnamefont {Vandersypen}},\ }\href {\doibase
  10.1126/science.1209524} {\bibfield  {journal} {\bibinfo  {journal}
  {Science}\ }\textbf {\bibinfo {volume} {333}},\ \bibinfo {pages} {1269}
  (\bibinfo {year} {2011})}\BibitemShut {NoStop}%
\bibitem [{\citenamefont {Botzem}\ \emph {et~al.}(2018)\citenamefont {Botzem},
  \citenamefont {Shulman}, \citenamefont {Foletti}, \citenamefont {Harvey},
  \citenamefont {Dial}, \citenamefont {Bethke}, \citenamefont {Cerfontaine},
  \citenamefont {McNeil}, \citenamefont {Mahalu}, \citenamefont {Umansky},
  \citenamefont {Ludwig}, \citenamefont {Wieck}, \citenamefont {Schuh},
  \citenamefont {Bougeard}, \citenamefont {Yacoby},\ and\ \citenamefont
  {Bluhm}}]{botzem_tuning_2018}%
  \BibitemOpen
  \bibfield  {author} {\bibinfo {author} {\bibfnamefont {T.}~\bibnamefont
  {Botzem}}, \bibinfo {author} {\bibfnamefont {M.~D.}\ \bibnamefont {Shulman}},
  \bibinfo {author} {\bibfnamefont {S.}~\bibnamefont {Foletti}}, \bibinfo
  {author} {\bibfnamefont {S.~P.}\ \bibnamefont {Harvey}}, \bibinfo {author}
  {\bibfnamefont {O.~E.}\ \bibnamefont {Dial}}, \bibinfo {author}
  {\bibfnamefont {P.}~\bibnamefont {Bethke}}, \bibinfo {author} {\bibfnamefont
  {P.}~\bibnamefont {Cerfontaine}}, \bibinfo {author} {\bibfnamefont
  {R.~P.~G.}\ \bibnamefont {McNeil}}, \bibinfo {author} {\bibfnamefont
  {D.}~\bibnamefont {Mahalu}}, \bibinfo {author} {\bibfnamefont
  {V.}~\bibnamefont {Umansky}}, \bibinfo {author} {\bibfnamefont
  {A.}~\bibnamefont {Ludwig}}, \bibinfo {author} {\bibfnamefont
  {A.}~\bibnamefont {Wieck}}, \bibinfo {author} {\bibfnamefont
  {D.}~\bibnamefont {Schuh}}, \bibinfo {author} {\bibfnamefont
  {D.}~\bibnamefont {Bougeard}}, \bibinfo {author} {\bibfnamefont
  {A.}~\bibnamefont {Yacoby}}, \ and\ \bibinfo {author} {\bibfnamefont
  {H.}~\bibnamefont {Bluhm}},\ }\href {\doibase
  10.1103/PhysRevApplied.10.054026} {\bibfield  {journal} {\bibinfo  {journal}
  {Physical Review Applied}\ }\textbf {\bibinfo {volume} {10}},\ \bibinfo
  {pages} {054026} (\bibinfo {year} {2018})}\BibitemShut {NoStop}%
\bibitem [{\citenamefont {West}\ \emph {et~al.}(2019)\citenamefont {West},
  \citenamefont {Hensen}, \citenamefont {Jouan}, \citenamefont {Tanttu},
  \citenamefont {Yang}, \citenamefont {Rossi}, \citenamefont {Gonzalez-Zalba},
  \citenamefont {Hudson}, \citenamefont {Morello}, \citenamefont {Reilly},\
  and\ \citenamefont {Dzurak}}]{west_gate-based_2019}%
  \BibitemOpen
  \bibfield  {author} {\bibinfo {author} {\bibfnamefont {A.}~\bibnamefont
  {West}}, \bibinfo {author} {\bibfnamefont {B.}~\bibnamefont {Hensen}},
  \bibinfo {author} {\bibfnamefont {A.}~\bibnamefont {Jouan}}, \bibinfo
  {author} {\bibfnamefont {T.}~\bibnamefont {Tanttu}}, \bibinfo {author}
  {\bibfnamefont {C.-H.}\ \bibnamefont {Yang}}, \bibinfo {author}
  {\bibfnamefont {A.}~\bibnamefont {Rossi}}, \bibinfo {author} {\bibfnamefont
  {M.~F.}\ \bibnamefont {Gonzalez-Zalba}}, \bibinfo {author} {\bibfnamefont
  {F.}~\bibnamefont {Hudson}}, \bibinfo {author} {\bibfnamefont
  {A.}~\bibnamefont {Morello}}, \bibinfo {author} {\bibfnamefont {D.~J.}\
  \bibnamefont {Reilly}}, \ and\ \bibinfo {author} {\bibfnamefont {A.~S.}\
  \bibnamefont {Dzurak}},\ }\href {\doibase 10.1038/s41565-019-0400-7}
  {\bibfield  {journal} {\bibinfo  {journal} {Nature Nanotechnology}\ }\textbf
  {\bibinfo {volume} {14}},\ \bibinfo {pages} {437} (\bibinfo {year}
  {2019})}\BibitemShut {NoStop}%
\bibitem [{\citenamefont {Takumi}\ \emph {et~al.}(2018)\citenamefont {Takumi},
  \citenamefont {Tomohiro}, \citenamefont {Takashi}, \citenamefont
  {Matthieu~R.}, \citenamefont {Shinichi}, \citenamefont {Jun}, \citenamefont
  {Kenta}, \citenamefont {Akito}, \citenamefont {Giles}, \citenamefont {Arne},
  \citenamefont {Andreas~D.},\ and\ \citenamefont
  {Seigo}}]{ito_four-single_spin_2018}%
  \BibitemOpen
  \bibfield  {author} {\bibinfo {author} {\bibfnamefont {I.}~\bibnamefont
  {Takumi}}, \bibinfo {author} {\bibfnamefont {O.}~\bibnamefont {Tomohiro}},
  \bibinfo {author} {\bibfnamefont {N.}~\bibnamefont {Takashi}}, \bibinfo
  {author} {\bibfnamefont {D.}~\bibnamefont {Matthieu~R.}}, \bibinfo {author}
  {\bibfnamefont {A.}~\bibnamefont {Shinichi}}, \bibinfo {author}
  {\bibfnamefont {Y.}~\bibnamefont {Jun}}, \bibinfo {author} {\bibfnamefont
  {T.}~\bibnamefont {Kenta}}, \bibinfo {author} {\bibfnamefont
  {N.}~\bibnamefont {Akito}}, \bibinfo {author} {\bibfnamefont
  {A.}~\bibnamefont {Giles}}, \bibinfo {author} {\bibfnamefont
  {L.}~\bibnamefont {Arne}}, \bibinfo {author} {\bibfnamefont {W.}~\bibnamefont
  {Andreas~D.}}, \ and\ \bibinfo {author} {\bibfnamefont {T.}~\bibnamefont
  {Seigo}},\ }\href {https://doi.org/10.1063/1.5040280} {\bibfield  {journal}
  {\bibinfo  {journal} {Appl. Phys. Lett.}\ }\textbf {\bibinfo {volume}
  {113}},\ \bibinfo {pages} {Appl. Phys. Lett.} (\bibinfo {year}
  {2018})}\BibitemShut {NoStop}%
\bibitem [{\citenamefont {Lim}\ \emph {et~al.}(2009)\citenamefont {Lim},
  \citenamefont {Zwanenburg}, \citenamefont {Huebl}, \citenamefont
  {M{\"o}tt{\"o}nen}, \citenamefont {Chan}, \citenamefont {Morello},\ and\
  \citenamefont {Dzurak}}]{lim_observation_2009}%
  \BibitemOpen
  \bibfield  {author} {\bibinfo {author} {\bibfnamefont {W.~H.}\ \bibnamefont
  {Lim}}, \bibinfo {author} {\bibfnamefont {F.~A.}\ \bibnamefont {Zwanenburg}},
  \bibinfo {author} {\bibfnamefont {H.}~\bibnamefont {Huebl}}, \bibinfo
  {author} {\bibfnamefont {M.}~\bibnamefont {M{\"o}tt{\"o}nen}}, \bibinfo
  {author} {\bibfnamefont {K.~W.}\ \bibnamefont {Chan}}, \bibinfo {author}
  {\bibfnamefont {A.}~\bibnamefont {Morello}}, \ and\ \bibinfo {author}
  {\bibfnamefont {A.~S.}\ \bibnamefont {Dzurak}},\ }\href {\doibase
  10.1063/1.3272858} {\bibfield  {journal} {\bibinfo  {journal} {Applied
  Physics Letters}\ }\textbf {\bibinfo {volume} {95}},\ \bibinfo {pages}
  {242102} (\bibinfo {year} {2009})}\BibitemShut {NoStop}%
\bibitem [{\citenamefont {Kouwen}\ \emph {et~al.}(2010)\citenamefont {Kouwen},
  \citenamefont {Reimer}, \citenamefont {Hidma}, \citenamefont {van Weert},
  \citenamefont {Algra}, \citenamefont {Bakkers}, \citenamefont {Kouwenhoven},\
  and\ \citenamefont {Zwiller}}]{kouwen_single_2010}%
  \BibitemOpen
  \bibfield  {author} {\bibinfo {author} {\bibfnamefont {M.~P.~v.}\
  \bibnamefont {Kouwen}}, \bibinfo {author} {\bibfnamefont {M.~E.}\
  \bibnamefont {Reimer}}, \bibinfo {author} {\bibfnamefont {A.~W.}\
  \bibnamefont {Hidma}}, \bibinfo {author} {\bibfnamefont {M.~H.~M.}\
  \bibnamefont {van Weert}}, \bibinfo {author} {\bibfnamefont {R.~E.}\
  \bibnamefont {Algra}}, \bibinfo {author} {\bibfnamefont {E.~P. A.~M.}\
  \bibnamefont {Bakkers}}, \bibinfo {author} {\bibfnamefont {L.~P.}\
  \bibnamefont {Kouwenhoven}}, \ and\ \bibinfo {author} {\bibfnamefont
  {V.}~\bibnamefont {Zwiller}},\ }\href {\doibase 10.1021/nl100520r} {\bibfield
   {journal} {\bibinfo  {journal} {Nano Letters}\ }\textbf {\bibinfo {volume}
  {10}},\ \bibinfo {pages} {1817} (\bibinfo {year} {2010})}\BibitemShut
  {NoStop}%
\bibitem [{\citenamefont {Mar}\ \emph {et~al.}(2011)\citenamefont {Mar},
  \citenamefont {Xu}, \citenamefont {Baumberg}, \citenamefont {Brossard},
  \citenamefont {Irvine}, \citenamefont {Stanley},\ and\ \citenamefont
  {Williams}}]{mar_bias-controlled_2011}%
  \BibitemOpen
  \bibfield  {author} {\bibinfo {author} {\bibfnamefont {J.~D.}\ \bibnamefont
  {Mar}}, \bibinfo {author} {\bibfnamefont {X.~L.}\ \bibnamefont {Xu}},
  \bibinfo {author} {\bibfnamefont {J.~J.}\ \bibnamefont {Baumberg}}, \bibinfo
  {author} {\bibfnamefont {F.~S.~F.}\ \bibnamefont {Brossard}}, \bibinfo
  {author} {\bibfnamefont {A.~C.}\ \bibnamefont {Irvine}}, \bibinfo {author}
  {\bibfnamefont {C.}~\bibnamefont {Stanley}}, \ and\ \bibinfo {author}
  {\bibfnamefont {D.~A.}\ \bibnamefont {Williams}},\ }\href {\doibase
  10.1103/PhysRevB.83.075306} {\bibfield  {journal} {\bibinfo  {journal}
  {Physical Review B}\ }\textbf {\bibinfo {volume} {83}},\ \bibinfo {pages}
  {075306} (\bibinfo {year} {2011})}\BibitemShut {NoStop}%
\bibitem [{\citenamefont {Kawakami}\ \emph {et~al.}(2014)\citenamefont
  {Kawakami}, \citenamefont {Scarlino}, \citenamefont {Ward}, \citenamefont
  {Braakman}, \citenamefont {Savage}, \citenamefont {Lagally}, \citenamefont
  {Friesen}, \citenamefont {Coppersmith}, \citenamefont {Eriksson},\ and\
  \citenamefont {Vandersypen}}]{kawakami_electrical_2014}%
  \BibitemOpen
  \bibfield  {author} {\bibinfo {author} {\bibfnamefont {E.}~\bibnamefont
  {Kawakami}}, \bibinfo {author} {\bibfnamefont {P.}~\bibnamefont {Scarlino}},
  \bibinfo {author} {\bibfnamefont {D.~R.}\ \bibnamefont {Ward}}, \bibinfo
  {author} {\bibfnamefont {F.~R.}\ \bibnamefont {Braakman}}, \bibinfo {author}
  {\bibfnamefont {D.~E.}\ \bibnamefont {Savage}}, \bibinfo {author}
  {\bibfnamefont {M.~G.}\ \bibnamefont {Lagally}}, \bibinfo {author}
  {\bibfnamefont {M.}~\bibnamefont {Friesen}}, \bibinfo {author} {\bibfnamefont
  {S.~N.}\ \bibnamefont {Coppersmith}}, \bibinfo {author} {\bibfnamefont
  {M.~A.}\ \bibnamefont {Eriksson}}, \ and\ \bibinfo {author} {\bibfnamefont
  {L.~M.~K.}\ \bibnamefont {Vandersypen}},\ }\href {\doibase
  10.1038/nnano.2014.153} {\bibfield  {journal} {\bibinfo  {journal} {Nature
  Nanotechnology}\ }\textbf {\bibinfo {volume} {9}},\ \bibinfo {pages} {666}
  (\bibinfo {year} {2014})}\BibitemShut {NoStop}%
\bibitem [{\citenamefont {Maurand}\ \emph {et~al.}(2016)\citenamefont
  {Maurand}, \citenamefont {Jehl}, \citenamefont {Kotekar-Patil}, \citenamefont
  {Corna}, \citenamefont {Bohuslavskyi}, \citenamefont {Lavi{\'e}ville},
  \citenamefont {Hutin}, \citenamefont {Barraud}, \citenamefont {Vinet},
  \citenamefont {Sanquer},\ and\ \citenamefont
  {De~Franceschi}}]{maurand_cmos_2016}%
  \BibitemOpen
  \bibfield  {author} {\bibinfo {author} {\bibfnamefont {R.}~\bibnamefont
  {Maurand}}, \bibinfo {author} {\bibfnamefont {X.}~\bibnamefont {Jehl}},
  \bibinfo {author} {\bibfnamefont {D.}~\bibnamefont {Kotekar-Patil}}, \bibinfo
  {author} {\bibfnamefont {A.}~\bibnamefont {Corna}}, \bibinfo {author}
  {\bibfnamefont {H.}~\bibnamefont {Bohuslavskyi}}, \bibinfo {author}
  {\bibfnamefont {R.}~\bibnamefont {Lavi{\'e}ville}}, \bibinfo {author}
  {\bibfnamefont {L.}~\bibnamefont {Hutin}}, \bibinfo {author} {\bibfnamefont
  {S.}~\bibnamefont {Barraud}}, \bibinfo {author} {\bibfnamefont
  {M.}~\bibnamefont {Vinet}}, \bibinfo {author} {\bibfnamefont
  {M.}~\bibnamefont {Sanquer}}, \ and\ \bibinfo {author} {\bibfnamefont
  {S.}~\bibnamefont {De~Franceschi}},\ }\href {\doibase 10.1038/ncomms13575}
  {\bibfield  {journal} {\bibinfo  {journal} {Nature Communications}\ }\textbf
  {\bibinfo {volume} {7}},\ \bibinfo {pages} {13575} (\bibinfo {year}
  {2016})}\BibitemShut {NoStop}%
\bibitem [{\citenamefont {Bayer}\ \emph {et~al.}(2000)\citenamefont {Bayer},
  \citenamefont {Stern}, \citenamefont {Hawrylak}, \citenamefont {Fafard},\
  and\ \citenamefont {Forchel}}]{bayer_hidden_2000}%
  \BibitemOpen
  \bibfield  {author} {\bibinfo {author} {\bibfnamefont {M.}~\bibnamefont
  {Bayer}}, \bibinfo {author} {\bibfnamefont {O.}~\bibnamefont {Stern}},
  \bibinfo {author} {\bibfnamefont {P.}~\bibnamefont {Hawrylak}}, \bibinfo
  {author} {\bibfnamefont {S.}~\bibnamefont {Fafard}}, \ and\ \bibinfo {author}
  {\bibfnamefont {A.}~\bibnamefont {Forchel}},\ }\href {\doibase
  10.1038/35016020} {\bibfield  {journal} {\bibinfo  {journal} {Nature}\
  }\textbf {\bibinfo {volume} {405}},\ \bibinfo {pages} {923} (\bibinfo {year}
  {2000})}\BibitemShut {NoStop}%
\bibitem [{\citenamefont {Hawrylak}\ and\ \citenamefont
  {Korkusinski}(2003)}]{michler_single_2003}%
  \BibitemOpen
  \bibfield  {author} {\bibinfo {author} {\bibfnamefont {P.}~\bibnamefont
  {Hawrylak}}\ and\ \bibinfo {author} {\bibfnamefont {M.}~\bibnamefont
  {Korkusinski}},\ }\href {\doibase 10.1007/b13751} {\emph {\bibinfo {title}
  {Chapter "Electronic Properties of Self-Assembled Quantum Dots" in "Single
  Quantum Dots - Fundamentals, Applications and New Concepts"}}}\ (\bibinfo
  {publisher} {Springer-Verlag Berlin Heidelberg},\ \bibinfo {year}
  {2003})\BibitemShut {NoStop}%
\bibitem [{\citenamefont {Raymond}\ \emph {et~al.}(2004)\citenamefont
  {Raymond}, \citenamefont {Studenikin}, \citenamefont {Sachrajda},
  \citenamefont {Wasilewski}, \citenamefont {Cheng}, \citenamefont {Sheng},
  \citenamefont {Hawrylak}, \citenamefont {Babinski}, \citenamefont {Potemski},
  \citenamefont {Ortner},\ and\ \citenamefont
  {Bayer}}]{raymond_excitonic_2004}%
  \BibitemOpen
  \bibfield  {author} {\bibinfo {author} {\bibfnamefont {S.}~\bibnamefont
  {Raymond}}, \bibinfo {author} {\bibfnamefont {S.}~\bibnamefont {Studenikin}},
  \bibinfo {author} {\bibfnamefont {A.}~\bibnamefont {Sachrajda}}, \bibinfo
  {author} {\bibfnamefont {Z.}~\bibnamefont {Wasilewski}}, \bibinfo {author}
  {\bibfnamefont {S.~J.}\ \bibnamefont {Cheng}}, \bibinfo {author}
  {\bibfnamefont {W.}~\bibnamefont {Sheng}}, \bibinfo {author} {\bibfnamefont
  {P.}~\bibnamefont {Hawrylak}}, \bibinfo {author} {\bibfnamefont
  {A.}~\bibnamefont {Babinski}}, \bibinfo {author} {\bibfnamefont
  {M.}~\bibnamefont {Potemski}}, \bibinfo {author} {\bibfnamefont
  {G.}~\bibnamefont {Ortner}}, \ and\ \bibinfo {author} {\bibfnamefont
  {M.}~\bibnamefont {Bayer}},\ }\href {\doibase 10.1103/PhysRevLett.92.187402}
  {\bibfield  {journal} {\bibinfo  {journal} {Physical Review Letters}\
  }\textbf {\bibinfo {volume} {92}},\ \bibinfo {pages} {187402} (\bibinfo
  {year} {2004})}\BibitemShut {NoStop}%
\bibitem [{\citenamefont {Castro~Neto}\ \emph {et~al.}(2009)\citenamefont
  {Castro~Neto}, \citenamefont {Guinea}, \citenamefont {Peres}, \citenamefont
  {Novoselov},\ and\ \citenamefont {Geim}}]{castro_neto_electronic_2009}%
  \BibitemOpen
  \bibfield  {author} {\bibinfo {author} {\bibfnamefont {A.~H.}\ \bibnamefont
  {Castro~Neto}}, \bibinfo {author} {\bibfnamefont {F.}~\bibnamefont {Guinea}},
  \bibinfo {author} {\bibfnamefont {N.~M.~R.}\ \bibnamefont {Peres}}, \bibinfo
  {author} {\bibfnamefont {K.~S.}\ \bibnamefont {Novoselov}}, \ and\ \bibinfo
  {author} {\bibfnamefont {A.~K.}\ \bibnamefont {Geim}},\ }\href {\doibase
  10.1103/RevModPhys.81.109} {\bibfield  {journal} {\bibinfo  {journal}
  {Reviews of Modern Physics}\ }\textbf {\bibinfo {volume} {81}},\ \bibinfo
  {pages} {109} (\bibinfo {year} {2009})}\BibitemShut {NoStop}%
\bibitem [{\citenamefont {Geim}\ and\ \citenamefont
  {Grigorieva}(2013)}]{geim_van_2013}%
  \BibitemOpen
  \bibfield  {author} {\bibinfo {author} {\bibfnamefont {A.~K.}\ \bibnamefont
  {Geim}}\ and\ \bibinfo {author} {\bibfnamefont {I.~V.}\ \bibnamefont
  {Grigorieva}},\ }\href {\doibase 10.1038/nature12385} {\bibfield  {journal}
  {\bibinfo  {journal} {Nature}\ }\textbf {\bibinfo {volume} {499}},\ \bibinfo
  {pages} {419} (\bibinfo {year} {2013})}\BibitemShut {NoStop}%
\bibitem [{\citenamefont {Ihn}\ \emph {et~al.}(2010)\citenamefont {Ihn},
  \citenamefont {G{\"u}ttinger}, \citenamefont {Molitor}, \citenamefont
  {Schnez}, \citenamefont {Schurtenberger}, \citenamefont {Jacobsen},
  \citenamefont {Hellm{\"u}ller}, \citenamefont {Frey}, \citenamefont
  {Dr{\"o}scher}, \citenamefont {Stampfer},\ and\ \citenamefont
  {Ensslin}}]{ihn_graphene_2010}%
  \BibitemOpen
  \bibfield  {author} {\bibinfo {author} {\bibfnamefont {T.}~\bibnamefont
  {Ihn}}, \bibinfo {author} {\bibfnamefont {J.}~\bibnamefont {G{\"u}ttinger}},
  \bibinfo {author} {\bibfnamefont {F.}~\bibnamefont {Molitor}}, \bibinfo
  {author} {\bibfnamefont {S.}~\bibnamefont {Schnez}}, \bibinfo {author}
  {\bibfnamefont {E.}~\bibnamefont {Schurtenberger}}, \bibinfo {author}
  {\bibfnamefont {A.}~\bibnamefont {Jacobsen}}, \bibinfo {author}
  {\bibfnamefont {S.}~\bibnamefont {Hellm{\"u}ller}}, \bibinfo {author}
  {\bibfnamefont {T.}~\bibnamefont {Frey}}, \bibinfo {author} {\bibfnamefont
  {S.}~\bibnamefont {Dr{\"o}scher}}, \bibinfo {author} {\bibfnamefont
  {C.}~\bibnamefont {Stampfer}}, \ and\ \bibinfo {author} {\bibfnamefont
  {K.}~\bibnamefont {Ensslin}},\ }\href {\doibase
  10.1016/S1369-7021(10)70033-X} {\bibfield  {journal} {\bibinfo  {journal}
  {Materials Today}\ }\textbf {\bibinfo {volume} {13}},\ \bibinfo {pages} {44}
  (\bibinfo {year} {2010})}\BibitemShut {NoStop}%
\bibitem [{\citenamefont {Mak}\ \emph {et~al.}(2010)\citenamefont {Mak},
  \citenamefont {Lee}, \citenamefont {Hone}, \citenamefont {Shan},\ and\
  \citenamefont {Heinz}}]{mak_atomically_2010}%
  \BibitemOpen
  \bibfield  {author} {\bibinfo {author} {\bibfnamefont {K.~F.}\ \bibnamefont
  {Mak}}, \bibinfo {author} {\bibfnamefont {C.}~\bibnamefont {Lee}}, \bibinfo
  {author} {\bibfnamefont {J.}~\bibnamefont {Hone}}, \bibinfo {author}
  {\bibfnamefont {J.}~\bibnamefont {Shan}}, \ and\ \bibinfo {author}
  {\bibfnamefont {T.~F.}\ \bibnamefont {Heinz}},\ }\href {\doibase
  10.1103/PhysRevLett.105.136805} {\bibfield  {journal} {\bibinfo  {journal}
  {Phys. Rev. Lett.}\ }\textbf {\bibinfo {volume} {105}},\ \bibinfo {pages}
  {136805} (\bibinfo {year} {2010})}\BibitemShut {NoStop}%
\bibitem [{\citenamefont {Manzeli}\ \emph {et~al.}(2017)\citenamefont
  {Manzeli}, \citenamefont {Ovchinnikov}, \citenamefont {Pasquier},
  \citenamefont {Yazyev},\ and\ \citenamefont {Kis}}]{manzeli_2d_2017}%
  \BibitemOpen
  \bibfield  {author} {\bibinfo {author} {\bibfnamefont {S.}~\bibnamefont
  {Manzeli}}, \bibinfo {author} {\bibfnamefont {D.}~\bibnamefont
  {Ovchinnikov}}, \bibinfo {author} {\bibfnamefont {D.}~\bibnamefont
  {Pasquier}}, \bibinfo {author} {\bibfnamefont {O.~V.}\ \bibnamefont
  {Yazyev}}, \ and\ \bibinfo {author} {\bibfnamefont {A.}~\bibnamefont {Kis}},\
  }\href {\doibase 10.1038/natrevmats.2017.33} {\bibfield  {journal} {\bibinfo
  {journal} {Nature Reviews Materials}\ }\textbf {\bibinfo {volume} {2}},\
  \bibinfo {pages} {17033} (\bibinfo {year} {2017})}\BibitemShut {NoStop}%
\bibitem [{\citenamefont {G{\"u}{\c c}l{\"u}}\ \emph
  {et~al.}(2014)\citenamefont {G{\"u}{\c c}l{\"u}}, \citenamefont {Potasz},
  \citenamefont {Korkusinski},\ and\ \citenamefont
  {Hawrylak}}]{guclu_graphene_2014}%
  \BibitemOpen
  \bibfield  {author} {\bibinfo {author} {\bibfnamefont {A.~D.}\ \bibnamefont
  {G{\"u}{\c c}l{\"u}}}, \bibinfo {author} {\bibfnamefont {P.}~\bibnamefont
  {Potasz}}, \bibinfo {author} {\bibfnamefont {M.}~\bibnamefont {Korkusinski}},
  \ and\ \bibinfo {author} {\bibfnamefont {P.}~\bibnamefont {Hawrylak}},\
  }\href@noop {} {\emph {\bibinfo {title} {Graphene quantum dots}}}\ (\bibinfo
  {publisher} {Springer},\ \bibinfo {year} {2014})\BibitemShut {NoStop}%
\bibitem [{\citenamefont {Scrace}\ \emph {et~al.}(2015)\citenamefont {Scrace},
  \citenamefont {Tsai}, \citenamefont {Barman}, \citenamefont {Schweidenback},
  \citenamefont {Petrou}, \citenamefont {Kioseoglou}, \citenamefont {Ozfidan},
  \citenamefont {Korkusinski},\ and\ \citenamefont
  {Hawrylak}}]{scrace_magnetoluminescence_2015}%
  \BibitemOpen
  \bibfield  {author} {\bibinfo {author} {\bibfnamefont {T.}~\bibnamefont
  {Scrace}}, \bibinfo {author} {\bibfnamefont {Y.}~\bibnamefont {Tsai}},
  \bibinfo {author} {\bibfnamefont {B.}~\bibnamefont {Barman}}, \bibinfo
  {author} {\bibfnamefont {L.}~\bibnamefont {Schweidenback}}, \bibinfo {author}
  {\bibfnamefont {A.}~\bibnamefont {Petrou}}, \bibinfo {author} {\bibfnamefont
  {G.}~\bibnamefont {Kioseoglou}}, \bibinfo {author} {\bibfnamefont
  {I.}~\bibnamefont {Ozfidan}}, \bibinfo {author} {\bibfnamefont
  {M.}~\bibnamefont {Korkusinski}}, \ and\ \bibinfo {author} {\bibfnamefont
  {P.}~\bibnamefont {Hawrylak}},\ }\href {\doibase 10.1038/nnano.2015.78}
  {\bibfield  {journal} {\bibinfo  {journal} {Nature Nanotechnology}\ }\textbf
  {\bibinfo {volume} {10}},\ \bibinfo {pages} {603} (\bibinfo {year}
  {2015})}\BibitemShut {NoStop}%
\bibitem [{\citenamefont {Kadantsev}\ and\ \citenamefont
  {Hawrylak}(2012)}]{kadantsev_electronic_2012}%
  \BibitemOpen
  \bibfield  {author} {\bibinfo {author} {\bibfnamefont {E.~S.}\ \bibnamefont
  {Kadantsev}}\ and\ \bibinfo {author} {\bibfnamefont {P.}~\bibnamefont
  {Hawrylak}},\ }\href {\doibase http://dx.doi.org/10.1016/j.ssc.2012.02.005}
  {\bibfield  {journal} {\bibinfo  {journal} {Solid State Communications}\
  }\textbf {\bibinfo {volume} {152}},\ \bibinfo {pages} {909 } (\bibinfo {year}
  {2012})}\BibitemShut {NoStop}%
\bibitem [{\citenamefont {Cao}\ \emph {et~al.}(2012)\citenamefont {Cao},
  \citenamefont {Wang}, \citenamefont {Han}, \citenamefont {Ye}, \citenamefont
  {Zhu}, \citenamefont {Shi}, \citenamefont {Niu}, \citenamefont {Tan},
  \citenamefont {Wang}, \citenamefont {Liu},\ and\ \citenamefont
  {Feng}}]{cao_valley-selective_2012}%
  \BibitemOpen
  \bibfield  {author} {\bibinfo {author} {\bibfnamefont {T.}~\bibnamefont
  {Cao}}, \bibinfo {author} {\bibfnamefont {G.}~\bibnamefont {Wang}}, \bibinfo
  {author} {\bibfnamefont {W.}~\bibnamefont {Han}}, \bibinfo {author}
  {\bibfnamefont {H.}~\bibnamefont {Ye}}, \bibinfo {author} {\bibfnamefont
  {C.}~\bibnamefont {Zhu}}, \bibinfo {author} {\bibfnamefont {J.}~\bibnamefont
  {Shi}}, \bibinfo {author} {\bibfnamefont {Q.}~\bibnamefont {Niu}}, \bibinfo
  {author} {\bibfnamefont {P.}~\bibnamefont {Tan}}, \bibinfo {author}
  {\bibfnamefont {E.}~\bibnamefont {Wang}}, \bibinfo {author} {\bibfnamefont
  {B.}~\bibnamefont {Liu}}, \ and\ \bibinfo {author} {\bibfnamefont
  {J.}~\bibnamefont {Feng}},\ }\href {\doibase 10.1038/ncomms1882} {\bibfield
  {journal} {\bibinfo  {journal} {Nature Communications}\ }\textbf {\bibinfo
  {volume} {3}},\ \bibinfo {pages} {887} (\bibinfo {year} {2012})}\BibitemShut
  {NoStop}%
\bibitem [{\citenamefont {Mak}\ \emph {et~al.}(2012)\citenamefont {Mak},
  \citenamefont {He}, \citenamefont {Shan},\ and\ \citenamefont
  {Heinz}}]{mak_control_2012}%
  \BibitemOpen
  \bibfield  {author} {\bibinfo {author} {\bibfnamefont {K.~F.}\ \bibnamefont
  {Mak}}, \bibinfo {author} {\bibfnamefont {K.}~\bibnamefont {He}}, \bibinfo
  {author} {\bibfnamefont {J.}~\bibnamefont {Shan}}, \ and\ \bibinfo {author}
  {\bibfnamefont {T.~F.}\ \bibnamefont {Heinz}},\ }\href {\doibase
  10.1038/nnano.2012.96} {\bibfield  {journal} {\bibinfo  {journal} {Nature
  Nanotechnology}\ }\textbf {\bibinfo {volume} {7}},\ \bibinfo {pages} {494}
  (\bibinfo {year} {2012})}\BibitemShut {NoStop}%
\bibitem [{\citenamefont {Jones}\ \emph {et~al.}(2013)\citenamefont {Jones},
  \citenamefont {Yu}, \citenamefont {Ghimire}, \citenamefont {Wu},
  \citenamefont {Aivazian}, \citenamefont {Ross}, \citenamefont {Zhao},
  \citenamefont {Yan}, \citenamefont {Mandrus}, \citenamefont {Xiao},\ and\
  \citenamefont {{others}}}]{jones_optical_2013}%
  \BibitemOpen
  \bibfield  {author} {\bibinfo {author} {\bibfnamefont {A.~M.}\ \bibnamefont
  {Jones}}, \bibinfo {author} {\bibfnamefont {H.}~\bibnamefont {Yu}}, \bibinfo
  {author} {\bibfnamefont {N.~J.}\ \bibnamefont {Ghimire}}, \bibinfo {author}
  {\bibfnamefont {S.}~\bibnamefont {Wu}}, \bibinfo {author} {\bibfnamefont
  {G.}~\bibnamefont {Aivazian}}, \bibinfo {author} {\bibfnamefont {J.~S.}\
  \bibnamefont {Ross}}, \bibinfo {author} {\bibfnamefont {B.}~\bibnamefont
  {Zhao}}, \bibinfo {author} {\bibfnamefont {J.}~\bibnamefont {Yan}}, \bibinfo
  {author} {\bibfnamefont {D.~G.}\ \bibnamefont {Mandrus}}, \bibinfo {author}
  {\bibfnamefont {D.}~\bibnamefont {Xiao}}, \ and\ \bibinfo {author}
  {\bibnamefont {{others}}},\ }\href {\doibase 10.1038/nnano.2013.151}
  {\bibfield  {journal} {\bibinfo  {journal} {Nat. Nano.}\ }\textbf {\bibinfo
  {volume} {8}},\ \bibinfo {pages} {634} (\bibinfo {year} {2013})}\BibitemShut
  {NoStop}%
\bibitem [{\citenamefont {Mak}\ \emph {et~al.}(2014)\citenamefont {Mak},
  \citenamefont {McGill}, \citenamefont {Park},\ and\ \citenamefont
  {McEuen}}]{mak_valley_2014}%
  \BibitemOpen
  \bibfield  {author} {\bibinfo {author} {\bibfnamefont {K.~F.}\ \bibnamefont
  {Mak}}, \bibinfo {author} {\bibfnamefont {K.~L.}\ \bibnamefont {McGill}},
  \bibinfo {author} {\bibfnamefont {J.}~\bibnamefont {Park}}, \ and\ \bibinfo
  {author} {\bibfnamefont {P.~L.}\ \bibnamefont {McEuen}},\ }\href {\doibase
  10.1126/science.1250140} {\bibfield  {journal} {\bibinfo  {journal}
  {Science}\ }\textbf {\bibinfo {volume} {344}},\ \bibinfo {pages} {1489}
  (\bibinfo {year} {2014})}\BibitemShut {NoStop}%
\bibitem [{\citenamefont {Szulakowska}\ \emph {et~al.}(2019)\citenamefont
  {Szulakowska}, \citenamefont {Bieniek},\ and\ \citenamefont
  {Hawrylak}}]{szulakowska_electronic_2019}%
  \BibitemOpen
  \bibfield  {author} {\bibinfo {author} {\bibfnamefont {L.}~\bibnamefont
  {Szulakowska}}, \bibinfo {author} {\bibfnamefont {M.}~\bibnamefont
  {Bieniek}}, \ and\ \bibinfo {author} {\bibfnamefont {P.}~\bibnamefont
  {Hawrylak}},\ }\href {\doibase 10.1016/j.sse.2019.03.002} {\bibfield
  {journal} {\bibinfo  {journal} {Solid-State Electronics}\ }\textbf {\bibinfo
  {volume} {155}},\ \bibinfo {pages} {105} (\bibinfo {year}
  {2019})}\BibitemShut {NoStop}%
\bibitem [{\citenamefont {Rose}\ \emph {et~al.}(2013)\citenamefont {Rose},
  \citenamefont {Goerbig},\ and\ \citenamefont
  {Pi{\'e}chon}}]{rose_spin-_2013}%
  \BibitemOpen
  \bibfield  {author} {\bibinfo {author} {\bibfnamefont {F.}~\bibnamefont
  {Rose}}, \bibinfo {author} {\bibfnamefont {M.~O.}\ \bibnamefont {Goerbig}}, \
  and\ \bibinfo {author} {\bibfnamefont {F.}~\bibnamefont {Pi{\'e}chon}},\
  }\href {\doibase 10.1103/PhysRevB.88.125438} {\bibfield  {journal} {\bibinfo
  {journal} {Physical Review B}\ }\textbf {\bibinfo {volume} {88}},\ \bibinfo
  {pages} {125438} (\bibinfo {year} {2013})}\BibitemShut {NoStop}%
\bibitem [{\citenamefont {Korm{\'a}nyos}\ \emph {et~al.}(2013)\citenamefont
  {Korm{\'a}nyos}, \citenamefont {Z{\'o}lyomi}, \citenamefont {Drummond},
  \citenamefont {Rakyta}, \citenamefont {Burkard},\ and\ \citenamefont
  {Fal'ko}}]{kormanyos_monolayer_2013}%
  \BibitemOpen
  \bibfield  {author} {\bibinfo {author} {\bibfnamefont {A.}~\bibnamefont
  {Korm{\'a}nyos}}, \bibinfo {author} {\bibfnamefont {V.}~\bibnamefont
  {Z{\'o}lyomi}}, \bibinfo {author} {\bibfnamefont {N.~D.}\ \bibnamefont
  {Drummond}}, \bibinfo {author} {\bibfnamefont {P.}~\bibnamefont {Rakyta}},
  \bibinfo {author} {\bibfnamefont {G.}~\bibnamefont {Burkard}}, \ and\
  \bibinfo {author} {\bibfnamefont {V.~I.}\ \bibnamefont {Fal'ko}},\ }\href
  {\doibase 10.1103/PhysRevB.88.045416} {\bibfield  {journal} {\bibinfo
  {journal} {Physical Review B}\ }\textbf {\bibinfo {volume} {88}},\ \bibinfo
  {pages} {045416} (\bibinfo {year} {2013})}\BibitemShut {NoStop}%
\bibitem [{\citenamefont {Paw{\l }owski}\ \emph {et~al.}(2018)\citenamefont
  {Paw{\l }owski}, \citenamefont {{\.Z}ebrowski},\ and\ \citenamefont
  {Bednarek}}]{pawlowski_valley_2018}%
  \BibitemOpen
  \bibfield  {author} {\bibinfo {author} {\bibfnamefont {J.}~\bibnamefont
  {Paw{\l }owski}}, \bibinfo {author} {\bibfnamefont {D.}~\bibnamefont
  {{\.Z}ebrowski}}, \ and\ \bibinfo {author} {\bibfnamefont {S.}~\bibnamefont
  {Bednarek}},\ }\href {\doibase 10.1103/PhysRevB.97.155412} {\bibfield
  {journal} {\bibinfo  {journal} {Physical Review B}\ }\textbf {\bibinfo
  {volume} {97}},\ \bibinfo {pages} {155412} (\bibinfo {year}
  {2018})}\BibitemShut {NoStop}%
\bibitem [{\citenamefont {Chirolli}\ \emph {et~al.}(2019)\citenamefont
  {Chirolli}, \citenamefont {Prada}, \citenamefont {Guinea}, \citenamefont
  {Rold{\'a}n},\ and\ \citenamefont {San-Jose}}]{chirolli_strain-induced_2019}%
  \BibitemOpen
  \bibfield  {author} {\bibinfo {author} {\bibfnamefont {L.}~\bibnamefont
  {Chirolli}}, \bibinfo {author} {\bibfnamefont {E.}~\bibnamefont {Prada}},
  \bibinfo {author} {\bibfnamefont {F.}~\bibnamefont {Guinea}}, \bibinfo
  {author} {\bibfnamefont {R.}~\bibnamefont {Rold{\'a}n}}, \ and\ \bibinfo
  {author} {\bibfnamefont {P.}~\bibnamefont {San-Jose}},\ }\href {\doibase
  10.1088/2053-1583/ab0113} {\bibfield  {journal} {\bibinfo  {journal} {2D
  Materials}\ }\textbf {\bibinfo {volume} {6}},\ \bibinfo {pages} {025010}
  (\bibinfo {year} {2019})}\BibitemShut {NoStop}%
\bibitem [{\citenamefont {Korm{\'a}nyos}\ \emph {et~al.}(2014)\citenamefont
  {Korm{\'a}nyos}, \citenamefont {Z{\'o}lyomi}, \citenamefont {Drummond},\ and\
  \citenamefont {Burkard}}]{kormanyos_spin-orbit_2014}%
  \BibitemOpen
  \bibfield  {author} {\bibinfo {author} {\bibfnamefont {A.}~\bibnamefont
  {Korm{\'a}nyos}}, \bibinfo {author} {\bibfnamefont {V.}~\bibnamefont
  {Z{\'o}lyomi}}, \bibinfo {author} {\bibfnamefont {N.~D.}\ \bibnamefont
  {Drummond}}, \ and\ \bibinfo {author} {\bibfnamefont {G.}~\bibnamefont
  {Burkard}},\ }\href {\doibase 10.1103/PhysRevX.4.011034} {\bibfield
  {journal} {\bibinfo  {journal} {Physical Review X}\ }\textbf {\bibinfo
  {volume} {4}},\ \bibinfo {pages} {011034} (\bibinfo {year}
  {2014})}\BibitemShut {NoStop}%
\bibitem [{\citenamefont {Liu}\ \emph {et~al.}(2014)\citenamefont {Liu},
  \citenamefont {Pang}, \citenamefont {Yao},\ and\ \citenamefont
  {Yao}}]{liu_intervalley_2014}%
  \BibitemOpen
  \bibfield  {author} {\bibinfo {author} {\bibfnamefont {G.-B.}\ \bibnamefont
  {Liu}}, \bibinfo {author} {\bibfnamefont {H.}~\bibnamefont {Pang}}, \bibinfo
  {author} {\bibfnamefont {Y.}~\bibnamefont {Yao}}, \ and\ \bibinfo {author}
  {\bibfnamefont {W.}~\bibnamefont {Yao}},\ }\href {\doibase
  10.1088/1367-2630/16/10/105011} {\bibfield  {journal} {\bibinfo  {journal}
  {New Journal of Physics}\ }\textbf {\bibinfo {volume} {16}},\ \bibinfo
  {pages} {105011} (\bibinfo {year} {2014})}\BibitemShut {NoStop}%
\bibitem [{\citenamefont {Dias}\ \emph {et~al.}(2016)\citenamefont {Dias},
  \citenamefont {Fu}, \citenamefont {Villegas-Lelovsky},\ and\ \citenamefont
  {Qu}}]{dias_robust_2016}%
  \BibitemOpen
  \bibfield  {author} {\bibinfo {author} {\bibfnamefont {A.~C.}\ \bibnamefont
  {Dias}}, \bibinfo {author} {\bibfnamefont {J.}~\bibnamefont {Fu}}, \bibinfo
  {author} {\bibfnamefont {L.}~\bibnamefont {Villegas-Lelovsky}}, \ and\
  \bibinfo {author} {\bibfnamefont {F.}~\bibnamefont {Qu}},\ }\href {\doibase
  10.1088/0953-8984/28/37/375803} {\bibfield  {journal} {\bibinfo  {journal}
  {Journal of Physics: Condensed Matter}\ }\textbf {\bibinfo {volume} {28}},\
  \bibinfo {pages} {375803} (\bibinfo {year} {2016})}\BibitemShut {NoStop}%
\bibitem [{\citenamefont {Wu}\ \emph {et~al.}(2016)\citenamefont {Wu},
  \citenamefont {Tong}, \citenamefont {Liu}, \citenamefont {Yu},\ and\
  \citenamefont {Yao}}]{wu_spin-valley_2016}%
  \BibitemOpen
  \bibfield  {author} {\bibinfo {author} {\bibfnamefont {Y.}~\bibnamefont
  {Wu}}, \bibinfo {author} {\bibfnamefont {Q.}~\bibnamefont {Tong}}, \bibinfo
  {author} {\bibfnamefont {G.-B.}\ \bibnamefont {Liu}}, \bibinfo {author}
  {\bibfnamefont {H.}~\bibnamefont {Yu}}, \ and\ \bibinfo {author}
  {\bibfnamefont {W.}~\bibnamefont {Yao}},\ }\href {\doibase
  10.1103/PhysRevB.93.045313} {\bibfield  {journal} {\bibinfo  {journal}
  {Physical Review B}\ }\textbf {\bibinfo {volume} {93}},\ \bibinfo {pages}
  {045313} (\bibinfo {year} {2016})}\BibitemShut {NoStop}%
\bibitem [{\citenamefont {Brooks}\ and\ \citenamefont
  {Burkard}(2017)}]{brooks_spin-degenerate_2017}%
  \BibitemOpen
  \bibfield  {author} {\bibinfo {author} {\bibfnamefont {M.}~\bibnamefont
  {Brooks}}\ and\ \bibinfo {author} {\bibfnamefont {G.}~\bibnamefont
  {Burkard}},\ }\href {\doibase 10.1103/PhysRevB.95.245411} {\bibfield
  {journal} {\bibinfo  {journal} {Physical Review B}\ }\textbf {\bibinfo
  {volume} {95}},\ \bibinfo {pages} {245411} (\bibinfo {year}
  {2017})}\BibitemShut {NoStop}%
\bibitem [{\citenamefont {Qu}\ \emph {et~al.}(2017)\citenamefont {Qu},
  \citenamefont {Dias}, \citenamefont {Fu}, \citenamefont {Villegas-Lelovsky},\
  and\ \citenamefont {Azevedo}}]{qu_tunable_2017}%
  \BibitemOpen
  \bibfield  {author} {\bibinfo {author} {\bibfnamefont {F.}~\bibnamefont
  {Qu}}, \bibinfo {author} {\bibfnamefont {A.~C.}\ \bibnamefont {Dias}},
  \bibinfo {author} {\bibfnamefont {J.}~\bibnamefont {Fu}}, \bibinfo {author}
  {\bibfnamefont {L.}~\bibnamefont {Villegas-Lelovsky}}, \ and\ \bibinfo
  {author} {\bibfnamefont {D.~L.}\ \bibnamefont {Azevedo}},\ }\href {\doibase
  10.1038/srep41044} {\bibfield  {journal} {\bibinfo  {journal} {Scientific
  Reports}\ }\textbf {\bibinfo {volume} {7}},\ \bibinfo {pages} {41044}
  (\bibinfo {year} {2017})}\BibitemShut {NoStop}%
\bibitem [{\citenamefont {Sz{\'e}chenyi}\ \emph {et~al.}(2018)\citenamefont
  {Sz{\'e}chenyi}, \citenamefont {Chirolli},\ and\ \citenamefont
  {P{\'a}lyi}}]{szechenyi_impurity-assisted_2018}%
  \BibitemOpen
  \bibfield  {author} {\bibinfo {author} {\bibfnamefont {G.}~\bibnamefont
  {Sz{\'e}chenyi}}, \bibinfo {author} {\bibfnamefont {L.}~\bibnamefont
  {Chirolli}}, \ and\ \bibinfo {author} {\bibfnamefont {A.}~\bibnamefont
  {P{\'a}lyi}},\ }\href {\doibase 10.1088/2053-1583/aab80e} {\bibfield
  {journal} {\bibinfo  {journal} {2D Materials}\ }\textbf {\bibinfo {volume}
  {5}},\ \bibinfo {pages} {035004} (\bibinfo {year} {2018})}\BibitemShut
  {NoStop}%
\bibitem [{\citenamefont {Pisoni}\ \emph {et~al.}(2018)\citenamefont {Pisoni},
  \citenamefont {Lei}, \citenamefont {Back}, \citenamefont {Eich},
  \citenamefont {Overweg}, \citenamefont {Lee}, \citenamefont {Watanabe},
  \citenamefont {Taniguchi}, \citenamefont {Ihn},\ and\ \citenamefont
  {Ensslin}}]{pisoni_gate-tunable_2018}%
  \BibitemOpen
  \bibfield  {author} {\bibinfo {author} {\bibfnamefont {R.}~\bibnamefont
  {Pisoni}}, \bibinfo {author} {\bibfnamefont {Z.}~\bibnamefont {Lei}},
  \bibinfo {author} {\bibfnamefont {P.}~\bibnamefont {Back}}, \bibinfo {author}
  {\bibfnamefont {M.}~\bibnamefont {Eich}}, \bibinfo {author} {\bibfnamefont
  {H.}~\bibnamefont {Overweg}}, \bibinfo {author} {\bibfnamefont
  {Y.}~\bibnamefont {Lee}}, \bibinfo {author} {\bibfnamefont {K.}~\bibnamefont
  {Watanabe}}, \bibinfo {author} {\bibfnamefont {T.}~\bibnamefont {Taniguchi}},
  \bibinfo {author} {\bibfnamefont {T.}~\bibnamefont {Ihn}}, \ and\ \bibinfo
  {author} {\bibfnamefont {K.}~\bibnamefont {Ensslin}},\ }\href {\doibase
  10.1063/1.5021113} {\bibfield  {journal} {\bibinfo  {journal} {Applied
  Physics Letters}\ }\textbf {\bibinfo {volume} {112}},\ \bibinfo {pages}
  {123101} (\bibinfo {year} {2018})}\BibitemShut {NoStop}%
\bibitem [{\citenamefont {Lu}\ \emph {et~al.}(2019)\citenamefont {Lu},
  \citenamefont {Chen}, \citenamefont {Dubey}, \citenamefont {Yao},
  \citenamefont {Li}, \citenamefont {Wang}, \citenamefont {Xiong},\ and\
  \citenamefont {Srivastava}}]{lu_optical_2019}%
  \BibitemOpen
  \bibfield  {author} {\bibinfo {author} {\bibfnamefont {X.}~\bibnamefont
  {Lu}}, \bibinfo {author} {\bibfnamefont {X.}~\bibnamefont {Chen}}, \bibinfo
  {author} {\bibfnamefont {S.}~\bibnamefont {Dubey}}, \bibinfo {author}
  {\bibfnamefont {Q.}~\bibnamefont {Yao}}, \bibinfo {author} {\bibfnamefont
  {W.}~\bibnamefont {Li}}, \bibinfo {author} {\bibfnamefont {X.}~\bibnamefont
  {Wang}}, \bibinfo {author} {\bibfnamefont {Q.}~\bibnamefont {Xiong}}, \ and\
  \bibinfo {author} {\bibfnamefont {A.}~\bibnamefont {Srivastava}},\ }\href
  {\doibase 10.1038/s41565-019-0394-1} {\bibfield  {journal} {\bibinfo
  {journal} {Nature Nanotechnology}\ }\textbf {\bibinfo {volume} {14}},\
  \bibinfo {pages} {426} (\bibinfo {year} {2019})}\BibitemShut {NoStop}%
\bibitem [{\citenamefont {Brotons-Gisbert}\ \emph {et~al.}(2019)\citenamefont
  {Brotons-Gisbert}, \citenamefont {Branny}, \citenamefont {Kumar},
  \citenamefont {Picard}, \citenamefont {Proux}, \citenamefont {Gray},
  \citenamefont {Burch}, \citenamefont {Watanabe}, \citenamefont {Taniguchi},\
  and\ \citenamefont {Gerardot}}]{brotons-gisbert_coulomb_2019}%
  \BibitemOpen
  \bibfield  {author} {\bibinfo {author} {\bibfnamefont {M.}~\bibnamefont
  {Brotons-Gisbert}}, \bibinfo {author} {\bibfnamefont {A.}~\bibnamefont
  {Branny}}, \bibinfo {author} {\bibfnamefont {S.}~\bibnamefont {Kumar}},
  \bibinfo {author} {\bibfnamefont {R.}~\bibnamefont {Picard}}, \bibinfo
  {author} {\bibfnamefont {R.}~\bibnamefont {Proux}}, \bibinfo {author}
  {\bibfnamefont {M.}~\bibnamefont {Gray}}, \bibinfo {author} {\bibfnamefont
  {K.~S.}\ \bibnamefont {Burch}}, \bibinfo {author} {\bibfnamefont
  {K.}~\bibnamefont {Watanabe}}, \bibinfo {author} {\bibfnamefont
  {T.}~\bibnamefont {Taniguchi}}, \ and\ \bibinfo {author} {\bibfnamefont
  {B.~D.}\ \bibnamefont {Gerardot}},\ }\href {\doibase
  10.1038/s41565-019-0402-5} {\bibfield  {journal} {\bibinfo  {journal} {Nature
  Nanotechnology}\ }\textbf {\bibinfo {volume} {14}},\ \bibinfo {pages} {442}
  (\bibinfo {year} {2019})}\BibitemShut {NoStop}%
\bibitem [{\citenamefont {Song}\ \emph {et~al.}(2015)\citenamefont {Song},
  \citenamefont {Liu}, \citenamefont {Mosallanejad}, \citenamefont {You},
  \citenamefont {Han}, \citenamefont {Chen}, \citenamefont {Li}, \citenamefont
  {Cao}, \citenamefont {Xiao}, \citenamefont {Guo},\ and\ \citenamefont
  {Guo}}]{song_gate_2015}%
  \BibitemOpen
  \bibfield  {author} {\bibinfo {author} {\bibfnamefont {X.-X.}\ \bibnamefont
  {Song}}, \bibinfo {author} {\bibfnamefont {D.}~\bibnamefont {Liu}}, \bibinfo
  {author} {\bibfnamefont {V.}~\bibnamefont {Mosallanejad}}, \bibinfo {author}
  {\bibfnamefont {J.}~\bibnamefont {You}}, \bibinfo {author} {\bibfnamefont
  {T.-Y.}\ \bibnamefont {Han}}, \bibinfo {author} {\bibfnamefont {D.-T.}\
  \bibnamefont {Chen}}, \bibinfo {author} {\bibfnamefont {H.-O.}\ \bibnamefont
  {Li}}, \bibinfo {author} {\bibfnamefont {G.}~\bibnamefont {Cao}}, \bibinfo
  {author} {\bibfnamefont {M.}~\bibnamefont {Xiao}}, \bibinfo {author}
  {\bibfnamefont {G.-C.}\ \bibnamefont {Guo}}, \ and\ \bibinfo {author}
  {\bibfnamefont {G.-P.}\ \bibnamefont {Guo}},\ }\href {\doibase
  10.1039/C5NR04961J} {\bibfield  {journal} {\bibinfo  {journal} {Nanoscale}\
  }\textbf {\bibinfo {volume} {7}},\ \bibinfo {pages} {16867} (\bibinfo {year}
  {2015})}\BibitemShut {NoStop}%
\bibitem [{\citenamefont {Zeng}\ \emph {et~al.}(2012)\citenamefont {Zeng},
  \citenamefont {Dai}, \citenamefont {Yao}, \citenamefont {Xiao},\ and\
  \citenamefont {Cui}}]{zeng_valley_2012}%
  \BibitemOpen
  \bibfield  {author} {\bibinfo {author} {\bibfnamefont {H.}~\bibnamefont
  {Zeng}}, \bibinfo {author} {\bibfnamefont {J.}~\bibnamefont {Dai}}, \bibinfo
  {author} {\bibfnamefont {W.}~\bibnamefont {Yao}}, \bibinfo {author}
  {\bibfnamefont {D.}~\bibnamefont {Xiao}}, \ and\ \bibinfo {author}
  {\bibfnamefont {X.}~\bibnamefont {Cui}},\ }\href {\doibase
  10.1038/nnano.2012.95} {\bibfield  {journal} {\bibinfo  {journal} {Nature
  Nanotechnology}\ }\textbf {\bibinfo {volume} {7}},\ \bibinfo {pages} {490}
  (\bibinfo {year} {2012})}\BibitemShut {NoStop}%
\bibitem [{\citenamefont {Bernardi}\ \emph {et~al.}(2013)\citenamefont
  {Bernardi}, \citenamefont {Palummo},\ and\ \citenamefont
  {Grossman}}]{bernardi_extraordinary_2013}%
  \BibitemOpen
  \bibfield  {author} {\bibinfo {author} {\bibfnamefont {M.}~\bibnamefont
  {Bernardi}}, \bibinfo {author} {\bibfnamefont {M.}~\bibnamefont {Palummo}}, \
  and\ \bibinfo {author} {\bibfnamefont {J.~C.}\ \bibnamefont {Grossman}},\
  }\href {\doibase 10.1021/nl401544y} {\bibfield  {journal} {\bibinfo
  {journal} {Nano Letters}\ }\textbf {\bibinfo {volume} {13}},\ \bibinfo
  {pages} {3664} (\bibinfo {year} {2013})}\BibitemShut {NoStop}%
\bibitem [{\citenamefont {Wang}\ \emph {et~al.}(2012)\citenamefont {Wang},
  \citenamefont {Kalantar-Zadeh}, \citenamefont {Kis}, \citenamefont
  {Coleman},\ and\ \citenamefont {Strano}}]{wang_electronics_2012}%
  \BibitemOpen
  \bibfield  {author} {\bibinfo {author} {\bibfnamefont {Q.~H.}\ \bibnamefont
  {Wang}}, \bibinfo {author} {\bibfnamefont {K.}~\bibnamefont
  {Kalantar-Zadeh}}, \bibinfo {author} {\bibfnamefont {A.}~\bibnamefont {Kis}},
  \bibinfo {author} {\bibfnamefont {J.~N.}\ \bibnamefont {Coleman}}, \ and\
  \bibinfo {author} {\bibfnamefont {M.~S.}\ \bibnamefont {Strano}},\ }\href
  {\doibase 10.1038/nnano.2012.193} {\bibfield  {journal} {\bibinfo  {journal}
  {Nature Nanotechnology}\ }\textbf {\bibinfo {volume} {7}},\ \bibinfo {pages}
  {699} (\bibinfo {year} {2012})}\BibitemShut {NoStop}%
\bibitem [{\citenamefont {Cong}\ \emph {et~al.}(2015)\citenamefont {Cong},
  \citenamefont {Tang}, \citenamefont {Zhao},\ and\ \citenamefont
  {Chu}}]{cong_enhanced_2015}%
  \BibitemOpen
  \bibfield  {author} {\bibinfo {author} {\bibfnamefont {W.~T.}\ \bibnamefont
  {Cong}}, \bibinfo {author} {\bibfnamefont {Z.}~\bibnamefont {Tang}}, \bibinfo
  {author} {\bibfnamefont {X.~G.}\ \bibnamefont {Zhao}}, \ and\ \bibinfo
  {author} {\bibfnamefont {J.~H.}\ \bibnamefont {Chu}},\ }\href {\doibase
  10.1038/srep09361} {\bibfield  {journal} {\bibinfo  {journal} {Scientific
  Reports}\ }\textbf {\bibinfo {volume} {5}},\ \bibinfo {pages} {9361}
  (\bibinfo {year} {2015})}\BibitemShut {NoStop}%
\bibitem [{\citenamefont {Mu}\ \emph {et~al.}(2018)\citenamefont {Mu},
  \citenamefont {Wei}, \citenamefont {Li}, \citenamefont {Huang},\ and\
  \citenamefont {Dai}}]{mu_electronic_2018}%
  \BibitemOpen
  \bibfield  {author} {\bibinfo {author} {\bibfnamefont {C.}~\bibnamefont
  {Mu}}, \bibinfo {author} {\bibfnamefont {W.}~\bibnamefont {Wei}}, \bibinfo
  {author} {\bibfnamefont {J.}~\bibnamefont {Li}}, \bibinfo {author}
  {\bibfnamefont {B.}~\bibnamefont {Huang}}, \ and\ \bibinfo {author}
  {\bibfnamefont {Y.}~\bibnamefont {Dai}},\ }\href {\doibase
  10.1088/2053-1591/aabddf} {\bibfield  {journal} {\bibinfo  {journal}
  {Materials Research Express}\ }\textbf {\bibinfo {volume} {5}},\ \bibinfo
  {pages} {046307} (\bibinfo {year} {2018})}\BibitemShut {NoStop}%
\bibitem [{\citenamefont {Miao}\ \emph {et~al.}(2018)\citenamefont {Miao},
  \citenamefont {Huang}, \citenamefont {Bao}, \citenamefont {Xu}, \citenamefont
  {Ma},\ and\ \citenamefont {Chu}}]{miao_tunable_2018}%
  \BibitemOpen
  \bibfield  {author} {\bibinfo {author} {\bibfnamefont {Y.}~\bibnamefont
  {Miao}}, \bibinfo {author} {\bibfnamefont {Y.}~\bibnamefont {Huang}},
  \bibinfo {author} {\bibfnamefont {H.}~\bibnamefont {Bao}}, \bibinfo {author}
  {\bibfnamefont {K.}~\bibnamefont {Xu}}, \bibinfo {author} {\bibfnamefont
  {F.}~\bibnamefont {Ma}}, \ and\ \bibinfo {author} {\bibfnamefont {P.~K.}\
  \bibnamefont {Chu}},\ }\href {\doibase 10.1088/1361-648X/aabd46} {\bibfield
  {journal} {\bibinfo  {journal} {Journal of Physics: Condensed Matter}\
  }\textbf {\bibinfo {volume} {30}},\ \bibinfo {pages} {215801} (\bibinfo
  {year} {2018})}\BibitemShut {NoStop}%
\bibitem [{\citenamefont {Frisenda}\ \emph {et~al.}(2017)\citenamefont
  {Frisenda}, \citenamefont {Dr{\"u}ppel}, \citenamefont {Schmidt},
  \citenamefont {Michaelis~de Vasconcellos}, \citenamefont {Perez~de Lara},
  \citenamefont {Bratschitsch}, \citenamefont {Rohlfing},\ and\ \citenamefont
  {Castellanos-Gomez}}]{frisenda_biaxial_2017}%
  \BibitemOpen
  \bibfield  {author} {\bibinfo {author} {\bibfnamefont {R.}~\bibnamefont
  {Frisenda}}, \bibinfo {author} {\bibfnamefont {M.}~\bibnamefont
  {Dr{\"u}ppel}}, \bibinfo {author} {\bibfnamefont {R.}~\bibnamefont
  {Schmidt}}, \bibinfo {author} {\bibfnamefont {S.}~\bibnamefont {Michaelis~de
  Vasconcellos}}, \bibinfo {author} {\bibfnamefont {D.}~\bibnamefont {Perez~de
  Lara}}, \bibinfo {author} {\bibfnamefont {R.}~\bibnamefont {Bratschitsch}},
  \bibinfo {author} {\bibfnamefont {M.}~\bibnamefont {Rohlfing}}, \ and\
  \bibinfo {author} {\bibfnamefont {A.}~\bibnamefont {Castellanos-Gomez}},\
  }\href {\doibase 10.1038/s41699-017-0013-7} {\bibfield  {journal} {\bibinfo
  {journal} {npj 2D Materials and Applications}\ }\textbf {\bibinfo {volume}
  {1}},\ \bibinfo {pages} {10} (\bibinfo {year} {2017})}\BibitemShut {NoStop}%
\bibitem [{\citenamefont {Yun}\ and\ \citenamefont
  {Lee}(2016)}]{yun_schottky_2016}%
  \BibitemOpen
  \bibfield  {author} {\bibinfo {author} {\bibfnamefont {W.~S.}\ \bibnamefont
  {Yun}}\ and\ \bibinfo {author} {\bibfnamefont {J.~D.}\ \bibnamefont {Lee}},\
  }\href {\doibase 10.1039/C6CP05384J} {\bibfield  {journal} {\bibinfo
  {journal} {Physical Chemistry Chemical Physics}\ }\textbf {\bibinfo {volume}
  {18}},\ \bibinfo {pages} {31027} (\bibinfo {year} {2016})}\BibitemShut
  {NoStop}%
\bibitem [{\citenamefont {Man}\ \emph {et~al.}(2016)\citenamefont {Man},
  \citenamefont {Deckoff-Jones}, \citenamefont {Winchester}, \citenamefont
  {Shi}, \citenamefont {Gupta}, \citenamefont {Mohite}, \citenamefont {Kar},
  \citenamefont {Kioupakis}, \citenamefont {Talapatra},\ and\ \citenamefont
  {Dani}}]{man_protecting_2016}%
  \BibitemOpen
  \bibfield  {author} {\bibinfo {author} {\bibfnamefont {M.~K.~L.}\
  \bibnamefont {Man}}, \bibinfo {author} {\bibfnamefont {S.}~\bibnamefont
  {Deckoff-Jones}}, \bibinfo {author} {\bibfnamefont {A.}~\bibnamefont
  {Winchester}}, \bibinfo {author} {\bibfnamefont {G.}~\bibnamefont {Shi}},
  \bibinfo {author} {\bibfnamefont {G.}~\bibnamefont {Gupta}}, \bibinfo
  {author} {\bibfnamefont {A.~D.}\ \bibnamefont {Mohite}}, \bibinfo {author}
  {\bibfnamefont {S.}~\bibnamefont {Kar}}, \bibinfo {author} {\bibfnamefont
  {E.}~\bibnamefont {Kioupakis}}, \bibinfo {author} {\bibfnamefont
  {S.}~\bibnamefont {Talapatra}}, \ and\ \bibinfo {author} {\bibfnamefont
  {K.~M.}\ \bibnamefont {Dani}},\ }\href {\doibase 10.1038/srep20890}
  {\bibfield  {journal} {\bibinfo  {journal} {Scientific Reports}\ }\textbf
  {\bibinfo {volume} {6}},\ \bibinfo {pages} {20890} (\bibinfo {year}
  {2016})}\BibitemShut {NoStop}%
\bibitem [{\citenamefont {Qu}\ \emph {et~al.}(2011)\citenamefont {Qu},
  \citenamefont {Santos}, \citenamefont {Azevedo}, \citenamefont
  {L{\'o}pez-Richard},\ and\ \citenamefont {Marques}}]{qu_tunable_2011}%
  \BibitemOpen
  \bibfield  {author} {\bibinfo {author} {\bibfnamefont {F.}~\bibnamefont
  {Qu}}, \bibinfo {author} {\bibfnamefont {D.~R.}\ \bibnamefont {Santos}},
  \bibinfo {author} {\bibfnamefont {R.~B.}\ \bibnamefont {Azevedo}}, \bibinfo
  {author} {\bibfnamefont {V.}~\bibnamefont {L{\'o}pez-Richard}}, \ and\
  \bibinfo {author} {\bibfnamefont {G.~E.}\ \bibnamefont {Marques}},\ }\href
  {\doibase 10.1088/1742-6596/334/1/012064} {\bibfield  {journal} {\bibinfo
  {journal} {Journal of Physics: Conference Series}\ }\textbf {\bibinfo
  {volume} {334}},\ \bibinfo {pages} {012064} (\bibinfo {year}
  {2011})}\BibitemShut {NoStop}%
\bibitem [{\citenamefont {Lee}\ \emph {et~al.}(2017)\citenamefont {Lee},
  \citenamefont {Wang}, \citenamefont {Xie}, \citenamefont {Mak},\ and\
  \citenamefont {Shan}}]{lee_valley_2017}%
  \BibitemOpen
  \bibfield  {author} {\bibinfo {author} {\bibfnamefont {J.}~\bibnamefont
  {Lee}}, \bibinfo {author} {\bibfnamefont {Z.}~\bibnamefont {Wang}}, \bibinfo
  {author} {\bibfnamefont {H.}~\bibnamefont {Xie}}, \bibinfo {author}
  {\bibfnamefont {K.~F.}\ \bibnamefont {Mak}}, \ and\ \bibinfo {author}
  {\bibfnamefont {J.}~\bibnamefont {Shan}},\ }\href {\doibase 10.1038/nmat4931}
  {\bibfield  {journal} {\bibinfo  {journal} {Nature Materials}\ }\textbf
  {\bibinfo {volume} {16}},\ \bibinfo {pages} {887} (\bibinfo {year}
  {2017})}\BibitemShut {NoStop}%
\bibitem [{\citenamefont {Wang}\ \emph
  {et~al.}(2017{\natexlab{a}})\citenamefont {Wang}, \citenamefont {Shan},\ and\
  \citenamefont {Mak}}]{wang_valley_2017}%
  \BibitemOpen
  \bibfield  {author} {\bibinfo {author} {\bibfnamefont {Z.}~\bibnamefont
  {Wang}}, \bibinfo {author} {\bibfnamefont {J.}~\bibnamefont {Shan}}, \ and\
  \bibinfo {author} {\bibfnamefont {K.~F.}\ \bibnamefont {Mak}},\ }\href
  {\doibase 10.1038/nnano.2016.213} {\bibfield  {journal} {\bibinfo  {journal}
  {Nature Nanotechnology}\ }\textbf {\bibinfo {volume} {12}},\ \bibinfo {pages}
  {144} (\bibinfo {year} {2017}{\natexlab{a}})}\BibitemShut {NoStop}%
\bibitem [{\citenamefont {Chen}\ \emph {et~al.}(2018)\citenamefont {Chen},
  \citenamefont {Li},\ and\ \citenamefont {Peeters}}]{chen_magnetic_2018}%
  \BibitemOpen
  \bibfield  {author} {\bibinfo {author} {\bibfnamefont {Q.}~\bibnamefont
  {Chen}}, \bibinfo {author} {\bibfnamefont {L.~L.}\ \bibnamefont {Li}}, \ and\
  \bibinfo {author} {\bibfnamefont {F.~M.}\ \bibnamefont {Peeters}},\ }\href
  {\doibase 10.1103/PhysRevB.97.085437} {\bibfield  {journal} {\bibinfo
  {journal} {Physical Review B}\ }\textbf {\bibinfo {volume} {97}},\ \bibinfo
  {pages} {085437} (\bibinfo {year} {2018})}\BibitemShut {NoStop}%
\bibitem [{\citenamefont {G{\"u}ttinger}\ \emph {et~al.}(2010)\citenamefont
  {G{\"u}ttinger}, \citenamefont {Frey}, \citenamefont {Stampfer},
  \citenamefont {Ihn},\ and\ \citenamefont {Ensslin}}]{guttinger_spin_2010}%
  \BibitemOpen
  \bibfield  {author} {\bibinfo {author} {\bibfnamefont {J.}~\bibnamefont
  {G{\"u}ttinger}}, \bibinfo {author} {\bibfnamefont {T.}~\bibnamefont {Frey}},
  \bibinfo {author} {\bibfnamefont {C.}~\bibnamefont {Stampfer}}, \bibinfo
  {author} {\bibfnamefont {T.}~\bibnamefont {Ihn}}, \ and\ \bibinfo {author}
  {\bibfnamefont {K.}~\bibnamefont {Ensslin}},\ }\href {\doibase
  10.1103/PhysRevLett.105.116801} {\bibfield  {journal} {\bibinfo  {journal}
  {Physical Review Letters}\ }\textbf {\bibinfo {volume} {105}},\ \bibinfo
  {pages} {116801} (\bibinfo {year} {2010})}\BibitemShut {NoStop}%
\bibitem [{\citenamefont {McGuire}(2016)}]{mcguire_growth_2016}%
  \BibitemOpen
  \bibfield  {author} {\bibinfo {author} {\bibfnamefont {J.~A.}\ \bibnamefont
  {McGuire}},\ }\href {\doibase 10.1002/pssr.201510287} {\bibfield  {journal}
  {\bibinfo  {journal} {Physica Status Solidi (RRL) - Rapid Research Letters}\
  }\textbf {\bibinfo {volume} {10}},\ \bibinfo {pages} {91} (\bibinfo {year}
  {2016})}\BibitemShut {NoStop}%
\bibitem [{\citenamefont {Wang}\ \emph
  {et~al.}(2017{\natexlab{b}})\citenamefont {Wang}, \citenamefont {Kharche},
  \citenamefont {Costa~Gir{\~a}o}, \citenamefont {Feng}, \citenamefont
  {M{\"u}llen}, \citenamefont {Meunier}, \citenamefont {Fasel},\ and\
  \citenamefont {Ruffieux}}]{wang_quantum_2017}%
  \BibitemOpen
  \bibfield  {author} {\bibinfo {author} {\bibfnamefont {S.}~\bibnamefont
  {Wang}}, \bibinfo {author} {\bibfnamefont {N.}~\bibnamefont {Kharche}},
  \bibinfo {author} {\bibfnamefont {E.}~\bibnamefont {Costa~Gir{\~a}o}},
  \bibinfo {author} {\bibfnamefont {X.}~\bibnamefont {Feng}}, \bibinfo {author}
  {\bibfnamefont {K.}~\bibnamefont {M{\"u}llen}}, \bibinfo {author}
  {\bibfnamefont {V.}~\bibnamefont {Meunier}}, \bibinfo {author} {\bibfnamefont
  {R.}~\bibnamefont {Fasel}}, \ and\ \bibinfo {author} {\bibfnamefont
  {P.}~\bibnamefont {Ruffieux}},\ }\href {\doibase
  10.1021/acs.nanolett.7b01244} {\bibfield  {journal} {\bibinfo  {journal}
  {Nano Letters}\ }\textbf {\bibinfo {volume} {17}},\ \bibinfo {pages} {4277}
  (\bibinfo {year} {2017}{\natexlab{b}})}\BibitemShut {NoStop}%
\bibitem [{\citenamefont {Wang}\ \emph {et~al.}(2018)\citenamefont {Wang},
  \citenamefont {De~Greve}, \citenamefont {Jauregui}, \citenamefont {Sushko},
  \citenamefont {High}, \citenamefont {Zhou}, \citenamefont {Scuri},
  \citenamefont {Taniguchi}, \citenamefont {Watanabe}, \citenamefont {Lukin},
  \citenamefont {Park},\ and\ \citenamefont {Kim}}]{wang_electrical_2018}%
  \BibitemOpen
  \bibfield  {author} {\bibinfo {author} {\bibfnamefont {K.}~\bibnamefont
  {Wang}}, \bibinfo {author} {\bibfnamefont {K.}~\bibnamefont {De~Greve}},
  \bibinfo {author} {\bibfnamefont {L.~A.}\ \bibnamefont {Jauregui}}, \bibinfo
  {author} {\bibfnamefont {A.}~\bibnamefont {Sushko}}, \bibinfo {author}
  {\bibfnamefont {A.}~\bibnamefont {High}}, \bibinfo {author} {\bibfnamefont
  {Y.}~\bibnamefont {Zhou}}, \bibinfo {author} {\bibfnamefont {G.}~\bibnamefont
  {Scuri}}, \bibinfo {author} {\bibfnamefont {T.}~\bibnamefont {Taniguchi}},
  \bibinfo {author} {\bibfnamefont {K.}~\bibnamefont {Watanabe}}, \bibinfo
  {author} {\bibfnamefont {M.~D.}\ \bibnamefont {Lukin}}, \bibinfo {author}
  {\bibfnamefont {H.}~\bibnamefont {Park}}, \ and\ \bibinfo {author}
  {\bibfnamefont {P.}~\bibnamefont {Kim}},\ }\href {\doibase
  10.1038/s41565-017-0030-x} {\bibfield  {journal} {\bibinfo  {journal} {Nature
  Nanotechnology}\ }\textbf {\bibinfo {volume} {13}},\ \bibinfo {pages} {128}
  (\bibinfo {year} {2018})}\BibitemShut {NoStop}%
\bibitem [{\citenamefont {Volk}\ \emph {et~al.}(2011)\citenamefont {Volk},
  \citenamefont {Fringes}, \citenamefont {Terr{\'e}s}, \citenamefont {Dauber},
  \citenamefont {Engels}, \citenamefont {Trellenkamp},\ and\ \citenamefont
  {Stampfer}}]{volk_electronic_2011}%
  \BibitemOpen
  \bibfield  {author} {\bibinfo {author} {\bibfnamefont {C.}~\bibnamefont
  {Volk}}, \bibinfo {author} {\bibfnamefont {S.}~\bibnamefont {Fringes}},
  \bibinfo {author} {\bibfnamefont {B.}~\bibnamefont {Terr{\'e}s}}, \bibinfo
  {author} {\bibfnamefont {J.}~\bibnamefont {Dauber}}, \bibinfo {author}
  {\bibfnamefont {S.}~\bibnamefont {Engels}}, \bibinfo {author} {\bibfnamefont
  {S.}~\bibnamefont {Trellenkamp}}, \ and\ \bibinfo {author} {\bibfnamefont
  {C.}~\bibnamefont {Stampfer}},\ }\href {\doibase 10.1021/nl201295s}
  {\bibfield  {journal} {\bibinfo  {journal} {Nano Letters}\ }\textbf {\bibinfo
  {volume} {11}},\ \bibinfo {pages} {3581} (\bibinfo {year}
  {2011})}\BibitemShut {NoStop}%
\bibitem [{\citenamefont {Allen}\ \emph {et~al.}(2012)\citenamefont {Allen},
  \citenamefont {Martin},\ and\ \citenamefont
  {Yacoby}}]{allen_gate-defined_2012}%
  \BibitemOpen
  \bibfield  {author} {\bibinfo {author} {\bibfnamefont {M.~T.}\ \bibnamefont
  {Allen}}, \bibinfo {author} {\bibfnamefont {J.}~\bibnamefont {Martin}}, \
  and\ \bibinfo {author} {\bibfnamefont {A.}~\bibnamefont {Yacoby}},\ }\href
  {\doibase 10.1038/ncomms1945} {\bibfield  {journal} {\bibinfo  {journal}
  {Nature Communications}\ }\textbf {\bibinfo {volume} {3}},\ \bibinfo {pages}
  {934} (\bibinfo {year} {2012})}\BibitemShut {NoStop}%
\bibitem [{\citenamefont {Eich}\ \emph {et~al.}(2018)\citenamefont {Eich},
  \citenamefont {Pisoni}, \citenamefont {Overweg}, \citenamefont {Kurzmann},
  \citenamefont {Lee}, \citenamefont {Rickhaus}, \citenamefont {Ihn},
  \citenamefont {Ensslin}, \citenamefont {Herman}, \citenamefont {Sigrist},
  \citenamefont {Watanabe},\ and\ \citenamefont {Taniguchi}}]{eich_spin_2018}%
  \BibitemOpen
  \bibfield  {author} {\bibinfo {author} {\bibfnamefont {M.}~\bibnamefont
  {Eich}}, \bibinfo {author} {\bibfnamefont {R.}~\bibnamefont {Pisoni}},
  \bibinfo {author} {\bibfnamefont {H.}~\bibnamefont {Overweg}}, \bibinfo
  {author} {\bibfnamefont {A.}~\bibnamefont {Kurzmann}}, \bibinfo {author}
  {\bibfnamefont {Y.}~\bibnamefont {Lee}}, \bibinfo {author} {\bibfnamefont
  {P.}~\bibnamefont {Rickhaus}}, \bibinfo {author} {\bibfnamefont
  {T.}~\bibnamefont {Ihn}}, \bibinfo {author} {\bibfnamefont {K.}~\bibnamefont
  {Ensslin}}, \bibinfo {author} {\bibfnamefont {F.~c.~v.}\ \bibnamefont
  {Herman}}, \bibinfo {author} {\bibfnamefont {M.}~\bibnamefont {Sigrist}},
  \bibinfo {author} {\bibfnamefont {K.}~\bibnamefont {Watanabe}}, \ and\
  \bibinfo {author} {\bibfnamefont {T.}~\bibnamefont {Taniguchi}},\ }\href
  {\doibase 10.1103/PhysRevX.8.031023} {\bibfield  {journal} {\bibinfo
  {journal} {Phys. Rev. X}\ }\textbf {\bibinfo {volume} {8}},\ \bibinfo {pages}
  {031023} (\bibinfo {year} {2018})}\BibitemShut {NoStop}%
\bibitem [{\citenamefont {Kurzmann}\ \emph {et~al.}(2019)\citenamefont
  {Kurzmann}, \citenamefont {Eich}, \citenamefont {Overweg}, \citenamefont
  {Mangold}, \citenamefont {Herman}, \citenamefont {Rickhaus}, \citenamefont
  {Pisoni}, \citenamefont {Lee}, \citenamefont {Garreis}, \citenamefont {Tong},
  \citenamefont {Watanabe}, \citenamefont {Taniguchi}, \citenamefont
  {Ensslin},\ and\ \citenamefont {Ihn}}]{kurzmann_ihn_2019}%
  \BibitemOpen
  \bibfield  {author} {\bibinfo {author} {\bibfnamefont {A.}~\bibnamefont
  {Kurzmann}}, \bibinfo {author} {\bibfnamefont {M.}~\bibnamefont {Eich}},
  \bibinfo {author} {\bibfnamefont {H.}~\bibnamefont {Overweg}}, \bibinfo
  {author} {\bibfnamefont {M.}~\bibnamefont {Mangold}}, \bibinfo {author}
  {\bibfnamefont {F.}~\bibnamefont {Herman}}, \bibinfo {author} {\bibfnamefont
  {P.}~\bibnamefont {Rickhaus}}, \bibinfo {author} {\bibfnamefont
  {R.}~\bibnamefont {Pisoni}}, \bibinfo {author} {\bibfnamefont
  {Y.}~\bibnamefont {Lee}}, \bibinfo {author} {\bibfnamefont {R.}~\bibnamefont
  {Garreis}}, \bibinfo {author} {\bibfnamefont {C.}~\bibnamefont {Tong}},
  \bibinfo {author} {\bibfnamefont {K.}~\bibnamefont {Watanabe}}, \bibinfo
  {author} {\bibfnamefont {T.}~\bibnamefont {Taniguchi}}, \bibinfo {author}
  {\bibfnamefont {K.}~\bibnamefont {Ensslin}}, \ and\ \bibinfo {author}
  {\bibfnamefont {T.}~\bibnamefont {Ihn}},\ }\href {\doibase
  10.1103/PhysRevLett.123.026803} {\bibfield  {journal} {\bibinfo  {journal}
  {Phys. Rev. Lett.}\ }\textbf {\bibinfo {volume} {123}},\ \bibinfo {pages}
  {026803} (\bibinfo {year} {2019})}\BibitemShut {NoStop}%
\bibitem [{\citenamefont {Huang}\ \emph {et~al.}(2014)\citenamefont {Huang},
  \citenamefont {Wu}, \citenamefont {Sanchez}, \citenamefont {Peters},
  \citenamefont {Beanland}, \citenamefont {Ross}, \citenamefont {Rivera},
  \citenamefont {Yao}, \citenamefont {Cobden},\ and\ \citenamefont
  {Xu}}]{huang_lateral_2014}%
  \BibitemOpen
  \bibfield  {author} {\bibinfo {author} {\bibfnamefont {C.}~\bibnamefont
  {Huang}}, \bibinfo {author} {\bibfnamefont {S.}~\bibnamefont {Wu}}, \bibinfo
  {author} {\bibfnamefont {A.~M.}\ \bibnamefont {Sanchez}}, \bibinfo {author}
  {\bibfnamefont {J.~J.~P.}\ \bibnamefont {Peters}}, \bibinfo {author}
  {\bibfnamefont {R.}~\bibnamefont {Beanland}}, \bibinfo {author}
  {\bibfnamefont {J.~S.}\ \bibnamefont {Ross}}, \bibinfo {author}
  {\bibfnamefont {P.}~\bibnamefont {Rivera}}, \bibinfo {author} {\bibfnamefont
  {W.}~\bibnamefont {Yao}}, \bibinfo {author} {\bibfnamefont {D.~H.}\
  \bibnamefont {Cobden}}, \ and\ \bibinfo {author} {\bibfnamefont
  {X.}~\bibnamefont {Xu}},\ }\href {\doibase 10.1038/nmat4064} {\bibfield
  {journal} {\bibinfo  {journal} {Nature Materials}\ }\textbf {\bibinfo
  {volume} {13}},\ \bibinfo {pages} {1096} (\bibinfo {year}
  {2014})}\BibitemShut {NoStop}%
\bibitem [{\citenamefont {Chakraborty}\ \emph {et~al.}(2018)\citenamefont
  {Chakraborty}, \citenamefont {Qiu}, \citenamefont {Konthasinghe},
  \citenamefont {Mukherjee}, \citenamefont {Dhara},\ and\ \citenamefont
  {Vamivakas}}]{chakraborty_3d_2018}%
  \BibitemOpen
  \bibfield  {author} {\bibinfo {author} {\bibfnamefont {C.}~\bibnamefont
  {Chakraborty}}, \bibinfo {author} {\bibfnamefont {L.}~\bibnamefont {Qiu}},
  \bibinfo {author} {\bibfnamefont {K.}~\bibnamefont {Konthasinghe}}, \bibinfo
  {author} {\bibfnamefont {A.}~\bibnamefont {Mukherjee}}, \bibinfo {author}
  {\bibfnamefont {S.}~\bibnamefont {Dhara}}, \ and\ \bibinfo {author}
  {\bibfnamefont {N.}~\bibnamefont {Vamivakas}},\ }\href {\doibase
  10.1021/acs.nanolett.7b05409} {\bibfield  {journal} {\bibinfo  {journal}
  {Nano Letters}\ }\textbf {\bibinfo {volume} {18}},\ \bibinfo {pages} {2859}
  (\bibinfo {year} {2018})}\BibitemShut {NoStop}%
\bibitem [{\citenamefont {Zhang}\ \emph {et~al.}(2017)\citenamefont {Zhang},
  \citenamefont {Song}, \citenamefont {Luo}, \citenamefont {Deng},
  \citenamefont {Mosallanejad}, \citenamefont {Taniguchi}, \citenamefont
  {Watanabe}, \citenamefont {Li}, \citenamefont {Cao}, \citenamefont {Guo},
  \citenamefont {Nori},\ and\ \citenamefont {Guo}}]{zhang_electrotunable_2017}%
  \BibitemOpen
  \bibfield  {author} {\bibinfo {author} {\bibfnamefont {Z.-Z.}\ \bibnamefont
  {Zhang}}, \bibinfo {author} {\bibfnamefont {X.-X.}\ \bibnamefont {Song}},
  \bibinfo {author} {\bibfnamefont {G.}~\bibnamefont {Luo}}, \bibinfo {author}
  {\bibfnamefont {G.-W.}\ \bibnamefont {Deng}}, \bibinfo {author}
  {\bibfnamefont {V.}~\bibnamefont {Mosallanejad}}, \bibinfo {author}
  {\bibfnamefont {T.}~\bibnamefont {Taniguchi}}, \bibinfo {author}
  {\bibfnamefont {K.}~\bibnamefont {Watanabe}}, \bibinfo {author}
  {\bibfnamefont {H.-O.}\ \bibnamefont {Li}}, \bibinfo {author} {\bibfnamefont
  {G.}~\bibnamefont {Cao}}, \bibinfo {author} {\bibfnamefont {G.-C.}\
  \bibnamefont {Guo}}, \bibinfo {author} {\bibfnamefont {F.}~\bibnamefont
  {Nori}}, \ and\ \bibinfo {author} {\bibfnamefont {G.-P.}\ \bibnamefont
  {Guo}},\ }\href {\doibase 10.1126/sciadv.1701699} {\bibfield  {journal}
  {\bibinfo  {journal} {Science Advances}\ }\textbf {\bibinfo {volume} {3}},\
  \bibinfo {pages} {e1701699} (\bibinfo {year} {2017})}\BibitemShut {NoStop}%
\bibitem [{\citenamefont {Pei}\ \emph {et~al.}(2015)\citenamefont {Pei},
  \citenamefont {Tao}, \citenamefont {Haibo},\ and\ \citenamefont
  {Song}}]{pei_structural_2015}%
  \BibitemOpen
  \bibfield  {author} {\bibinfo {author} {\bibfnamefont {L.}~\bibnamefont
  {Pei}}, \bibinfo {author} {\bibfnamefont {S.}~\bibnamefont {Tao}}, \bibinfo
  {author} {\bibfnamefont {S.}~\bibnamefont {Haibo}}, \ and\ \bibinfo {author}
  {\bibfnamefont {X.}~\bibnamefont {Song}},\ }\href {\doibase
  10.1016/j.ssc.2015.06.008} {\bibfield  {journal} {\bibinfo  {journal} {Solid
  State Communications}\ }\textbf {\bibinfo {volume} {218}},\ \bibinfo {pages}
  {25} (\bibinfo {year} {2015})}\BibitemShut {NoStop}%
\bibitem [{\citenamefont {Javaid}\ \emph {et~al.}(2017)\citenamefont {Javaid},
  \citenamefont {Drumm}, \citenamefont {Russo},\ and\ \citenamefont
  {Greentree}}]{javaid_study_2017}%
  \BibitemOpen
  \bibfield  {author} {\bibinfo {author} {\bibfnamefont {M.}~\bibnamefont
  {Javaid}}, \bibinfo {author} {\bibfnamefont {D.~W.}\ \bibnamefont {Drumm}},
  \bibinfo {author} {\bibfnamefont {S.~P.}\ \bibnamefont {Russo}}, \ and\
  \bibinfo {author} {\bibfnamefont {A.~D.}\ \bibnamefont {Greentree}},\ }\href
  {\doibase 10.1038/s41598-017-09305-y} {\bibfield  {journal} {\bibinfo
  {journal} {Scientific Reports}\ }\textbf {\bibinfo {volume} {7}} (\bibinfo
  {year} {2017}),\ 10.1038/s41598-017-09305-y}\BibitemShut {NoStop}%
\bibitem [{\citenamefont {Lauritsen}\ \emph {et~al.}(2003)\citenamefont
  {Lauritsen}, \citenamefont {Nyberg}, \citenamefont {Vang}, \citenamefont
  {Bollinger}, \citenamefont {Clausen}, \citenamefont {Topse}, \citenamefont
  {Jacobsen}, \citenamefont {L{\ae}gsgaard}, \citenamefont {N{\o}rskov},\ and\
  \citenamefont {Besenbacher}}]{lauritsen_chemistry_2003}%
  \BibitemOpen
  \bibfield  {author} {\bibinfo {author} {\bibfnamefont {J.~V.}\ \bibnamefont
  {Lauritsen}}, \bibinfo {author} {\bibfnamefont {M.}~\bibnamefont {Nyberg}},
  \bibinfo {author} {\bibfnamefont {R.~T.}\ \bibnamefont {Vang}}, \bibinfo
  {author} {\bibfnamefont {M.~V.}\ \bibnamefont {Bollinger}}, \bibinfo {author}
  {\bibfnamefont {B.~S.}\ \bibnamefont {Clausen}}, \bibinfo {author}
  {\bibfnamefont {H.}~\bibnamefont {Topse}}, \bibinfo {author} {\bibfnamefont
  {K.~W.}\ \bibnamefont {Jacobsen}}, \bibinfo {author} {\bibfnamefont
  {E.}~\bibnamefont {L{\ae}gsgaard}}, \bibinfo {author} {\bibfnamefont {J.~K.}\
  \bibnamefont {N{\o}rskov}}, \ and\ \bibinfo {author} {\bibfnamefont
  {F.}~\bibnamefont {Besenbacher}},\ }\href {\doibase
  10.1088/0957-4484/14/3/306} {\bibfield  {journal} {\bibinfo  {journal}
  {Nanotechnology}\ }\textbf {\bibinfo {volume} {14}},\ \bibinfo {pages} {385}
  (\bibinfo {year} {2003})}\BibitemShut {NoStop}%
\bibitem [{\citenamefont {Lauritsen}\ \emph {et~al.}(2007)\citenamefont
  {Lauritsen}, \citenamefont {Kibsgaard}, \citenamefont {Helveg}, \citenamefont
  {Tops{\o}e}, \citenamefont {Clausen}, \citenamefont {L{\ae}gsgaard},\ and\
  \citenamefont {Besenbacher}}]{lauritsen_size-dependent_2007}%
  \BibitemOpen
  \bibfield  {author} {\bibinfo {author} {\bibfnamefont {J.~V.}\ \bibnamefont
  {Lauritsen}}, \bibinfo {author} {\bibfnamefont {J.}~\bibnamefont
  {Kibsgaard}}, \bibinfo {author} {\bibfnamefont {S.}~\bibnamefont {Helveg}},
  \bibinfo {author} {\bibfnamefont {H.}~\bibnamefont {Tops{\o}e}}, \bibinfo
  {author} {\bibfnamefont {B.~S.}\ \bibnamefont {Clausen}}, \bibinfo {author}
  {\bibfnamefont {E.}~\bibnamefont {L{\ae}gsgaard}}, \ and\ \bibinfo {author}
  {\bibfnamefont {F.}~\bibnamefont {Besenbacher}},\ }\href {\doibase
  10.1038/nnano.2006.171} {\bibfield  {journal} {\bibinfo  {journal} {Nature
  Nanotechnology}\ }\textbf {\bibinfo {volume} {2}},\ \bibinfo {pages} {53}
  (\bibinfo {year} {2007})}\BibitemShut {NoStop}%
\bibitem [{\citenamefont {McBride}\ and\ \citenamefont
  {Head}(2009)}]{mcbride_dft_2009}%
  \BibitemOpen
  \bibfield  {author} {\bibinfo {author} {\bibfnamefont {K.~L.}\ \bibnamefont
  {McBride}}\ and\ \bibinfo {author} {\bibfnamefont {J.~D.}\ \bibnamefont
  {Head}},\ }\href {\doibase 10.1002/qua.22328} {\bibfield  {journal} {\bibinfo
   {journal} {International Journal of Quantum Chemistry}\ }\textbf {\bibinfo
  {volume} {109}},\ \bibinfo {pages} {3570} (\bibinfo {year}
  {2009})}\BibitemShut {NoStop}%
\bibitem [{\citenamefont {Li}\ and\ \citenamefont
  {Galli}(2007)}]{li_electronic_2007}%
  \BibitemOpen
  \bibfield  {author} {\bibinfo {author} {\bibfnamefont {T.}~\bibnamefont
  {Li}}\ and\ \bibinfo {author} {\bibfnamefont {G.}~\bibnamefont {Galli}},\
  }\href {\doibase 10.1021/jp075424v} {\bibfield  {journal} {\bibinfo
  {journal} {The Journal of Physical Chemistry C}\ }\textbf {\bibinfo {volume}
  {111}},\ \bibinfo {pages} {16192} (\bibinfo {year} {2007})}\BibitemShut
  {NoStop}%
\bibitem [{\citenamefont {Pavlovi{\'c}}\ and\ \citenamefont
  {Peeters}(2015)}]{pavlovic_electronic_2015}%
  \BibitemOpen
  \bibfield  {author} {\bibinfo {author} {\bibfnamefont {S.}~\bibnamefont
  {Pavlovi{\'c}}}\ and\ \bibinfo {author} {\bibfnamefont {F.~M.}\ \bibnamefont
  {Peeters}},\ }\href {\doibase 10.1103/PhysRevB.91.155410} {\bibfield
  {journal} {\bibinfo  {journal} {Physical Review B}\ }\textbf {\bibinfo
  {volume} {91}},\ \bibinfo {pages} {155410} (\bibinfo {year}
  {2015})}\BibitemShut {NoStop}%
\bibitem [{\citenamefont {Zebrowski}\ \emph {et~al.}(2013)\citenamefont
  {Zebrowski}, \citenamefont {Wach},\ and\ \citenamefont
  {Szafran}}]{Zebrowski_Szafran_2013}%
  \BibitemOpen
  \bibfield  {author} {\bibinfo {author} {\bibfnamefont {D.~P.}\ \bibnamefont
  {Zebrowski}}, \bibinfo {author} {\bibfnamefont {E.}~\bibnamefont {Wach}}, \
  and\ \bibinfo {author} {\bibfnamefont {B.}~\bibnamefont {Szafran}},\ }\href
  {\doibase 10.1103/PhysRevB.88.165405} {\bibfield  {journal} {\bibinfo
  {journal} {Phys. Rev. B}\ }\textbf {\bibinfo {volume} {88}},\ \bibinfo
  {pages} {165405} (\bibinfo {year} {2013})}\BibitemShut {NoStop}%
\bibitem [{\citenamefont {Szafran}\ \emph {et~al.}(2018)\citenamefont
  {Szafran}, \citenamefont {Zebrowski},\ and\ \citenamefont
  {Mrenca-Kolasinska}}]{Szafran_Kolasinska_2018}%
  \BibitemOpen
  \bibfield  {author} {\bibinfo {author} {\bibfnamefont {B.}~\bibnamefont
  {Szafran}}, \bibinfo {author} {\bibfnamefont {D.}~\bibnamefont {Zebrowski}},
  \ and\ \bibinfo {author} {\bibfnamefont {A.}~\bibnamefont
  {Mrenca-Kolasinska}},\ }\href {\doibase 10.1038/s41598-018-25534-1}
  {\bibfield  {journal} {\bibinfo  {journal} {Scientific Reports}\ }\textbf
  {\bibinfo {volume} {8}},\ \bibinfo {pages} {7166} (\bibinfo {year}
  {2018})}\BibitemShut {NoStop}%
\bibitem [{\citenamefont {Szafran}\ and\ \citenamefont
  {Zebrowski}(2018)}]{Szafran_Zebrowski_2018}%
  \BibitemOpen
  \bibfield  {author} {\bibinfo {author} {\bibfnamefont {B.}~\bibnamefont
  {Szafran}}\ and\ \bibinfo {author} {\bibfnamefont {D.}~\bibnamefont
  {Zebrowski}},\ }\href {\doibase 10.1103/PhysRevB.98.155305} {\bibfield
  {journal} {\bibinfo  {journal} {Phys. Rev. B}\ }\textbf {\bibinfo {volume}
  {98}},\ \bibinfo {pages} {155305} (\bibinfo {year} {2018})}\BibitemShut
  {NoStop}%
\bibitem [{\citenamefont {David}\ \emph {et~al.}(2018)\citenamefont {David},
  \citenamefont {Burkard},\ and\ \citenamefont
  {Korm{\'a}nyos}}]{david_effective_2018}%
  \BibitemOpen
  \bibfield  {author} {\bibinfo {author} {\bibfnamefont {A.}~\bibnamefont
  {David}}, \bibinfo {author} {\bibfnamefont {G.}~\bibnamefont {Burkard}}, \
  and\ \bibinfo {author} {\bibfnamefont {A.}~\bibnamefont {Korm{\'a}nyos}},\
  }\href {\doibase 10.1088/2053-1583/aac17f} {\bibfield  {journal} {\bibinfo
  {journal} {2D Materials}\ }\textbf {\bibinfo {volume} {5}},\ \bibinfo {pages}
  {035031} (\bibinfo {year} {2018})}\BibitemShut {NoStop}%
\bibitem [{\citenamefont {Bieniek}\ \emph {et~al.}(2018)\citenamefont
  {Bieniek}, \citenamefont {Korkusi\ifmmode~\acute{n}\else \'{n}\fi{}ski},
  \citenamefont {Szulakowska}, \citenamefont {Potasz}, \citenamefont
  {Ozfidan},\ and\ \citenamefont {Hawrylak}}]{bieniek_band_2018}%
  \BibitemOpen
  \bibfield  {author} {\bibinfo {author} {\bibfnamefont {M.}~\bibnamefont
  {Bieniek}}, \bibinfo {author} {\bibfnamefont {M.}~\bibnamefont
  {Korkusi\ifmmode~\acute{n}\else \'{n}\fi{}ski}}, \bibinfo {author}
  {\bibfnamefont {L.}~\bibnamefont {Szulakowska}}, \bibinfo {author}
  {\bibfnamefont {P.}~\bibnamefont {Potasz}}, \bibinfo {author} {\bibfnamefont
  {I.}~\bibnamefont {Ozfidan}}, \ and\ \bibinfo {author} {\bibfnamefont
  {P.}~\bibnamefont {Hawrylak}},\ }\href {\doibase 10.1103/PhysRevB.97.085153}
  {\bibfield  {journal} {\bibinfo  {journal} {Phys. Rev. B}\ }\textbf {\bibinfo
  {volume} {97}},\ \bibinfo {pages} {085153} (\bibinfo {year}
  {2018})}\BibitemShut {NoStop}%
\bibitem [{\citenamefont {Wu}\ \emph {et~al.}(2015)\citenamefont {Wu},
  \citenamefont {Qu},\ and\ \citenamefont {MacDonald}}]{wu_exciton_2015}%
  \BibitemOpen
  \bibfield  {author} {\bibinfo {author} {\bibfnamefont {F.}~\bibnamefont
  {Wu}}, \bibinfo {author} {\bibfnamefont {F.}~\bibnamefont {Qu}}, \ and\
  \bibinfo {author} {\bibfnamefont {A.~H.}\ \bibnamefont {MacDonald}},\ }\href
  {\doibase 10.1103/PhysRevB.91.075310} {\bibfield  {journal} {\bibinfo
  {journal} {Phys. Rev. B}\ }\textbf {\bibinfo {volume} {91}},\ \bibinfo
  {pages} {075310} (\bibinfo {year} {2015})}\BibitemShut {NoStop}%
\bibitem [{\citenamefont {Zhou}\ \emph {et~al.}(2015)\citenamefont {Zhou},
  \citenamefont {Shan}, \citenamefont {Yao},\ and\ \citenamefont
  {Xiao}}]{zhou_berry_2015}%
  \BibitemOpen
  \bibfield  {author} {\bibinfo {author} {\bibfnamefont {J.}~\bibnamefont
  {Zhou}}, \bibinfo {author} {\bibfnamefont {W.-Y.}\ \bibnamefont {Shan}},
  \bibinfo {author} {\bibfnamefont {W.}~\bibnamefont {Yao}}, \ and\ \bibinfo
  {author} {\bibfnamefont {D.}~\bibnamefont {Xiao}},\ }\href {\doibase
  10.1103/PhysRevLett.115.166803} {\bibfield  {journal} {\bibinfo  {journal}
  {Phys. Rev. Lett.}\ }\textbf {\bibinfo {volume} {115}},\ \bibinfo {pages}
  {166803} (\bibinfo {year} {2015})}\BibitemShut {NoStop}%
\bibitem [{\citenamefont {Srivastava}\ and\ \citenamefont
  {Imamoglu}(2015)}]{srivastava_signatures_2015}%
  \BibitemOpen
  \bibfield  {author} {\bibinfo {author} {\bibfnamefont {A.}~\bibnamefont
  {Srivastava}}\ and\ \bibinfo {author} {\bibfnamefont {A.}~\bibnamefont
  {Imamoglu}},\ }\href {\doibase 10.1103/PhysRevLett.115.166802} {\bibfield
  {journal} {\bibinfo  {journal} {Phys. Rev. Lett.}\ }\textbf {\bibinfo
  {volume} {115}},\ \bibinfo {pages} {166802} (\bibinfo {year}
  {2015})}\BibitemShut {NoStop}%
\bibitem [{\citenamefont {Bao}\ \emph {et~al.}(2019)\citenamefont {Bao},
  \citenamefont {Cheung},\ and\ \citenamefont {Zhang}}]{bao_flavor_2019}%
  \BibitemOpen
  \bibfield  {author} {\bibinfo {author} {\bibfnamefont {Z.-q.}\ \bibnamefont
  {Bao}}, \bibinfo {author} {\bibfnamefont {P.}~\bibnamefont {Cheung}}, \ and\
  \bibinfo {author} {\bibfnamefont {F.}~\bibnamefont {Zhang}},\ }\href
  {http://arxiv.org/abs/1903.01967} {\bibfield  {journal} {\bibinfo  {journal}
  {arXiv:1903.01967 [cond-mat]}\ } (\bibinfo {year} {2019})},\ \bibinfo {note}
  {arXiv: 1903.01967}\BibitemShut {NoStop}%
\bibitem [{\citenamefont {Zhang}\ \emph {et~al.}(2014)\citenamefont {Zhang},
  \citenamefont {Chang}, \citenamefont {Zhou}, \citenamefont {Cui},
  \citenamefont {Yan}, \citenamefont {Liu}, \citenamefont {Schmitt},
  \citenamefont {Lee}, \citenamefont {Moore}, \citenamefont {Chen},
  \citenamefont {Lin}, \citenamefont {Jeng}, \citenamefont {Mo}, \citenamefont
  {Hussain}, \citenamefont {Bansil},\ and\ \citenamefont
  {Shen}}]{zhang_direct_2014}%
  \BibitemOpen
  \bibfield  {author} {\bibinfo {author} {\bibfnamefont {Y.}~\bibnamefont
  {Zhang}}, \bibinfo {author} {\bibfnamefont {T.-R.}\ \bibnamefont {Chang}},
  \bibinfo {author} {\bibfnamefont {B.}~\bibnamefont {Zhou}}, \bibinfo {author}
  {\bibfnamefont {Y.-T.}\ \bibnamefont {Cui}}, \bibinfo {author} {\bibfnamefont
  {H.}~\bibnamefont {Yan}}, \bibinfo {author} {\bibfnamefont {Z.}~\bibnamefont
  {Liu}}, \bibinfo {author} {\bibfnamefont {F.}~\bibnamefont {Schmitt}},
  \bibinfo {author} {\bibfnamefont {J.}~\bibnamefont {Lee}}, \bibinfo {author}
  {\bibfnamefont {R.}~\bibnamefont {Moore}}, \bibinfo {author} {\bibfnamefont
  {Y.}~\bibnamefont {Chen}}, \bibinfo {author} {\bibfnamefont {H.}~\bibnamefont
  {Lin}}, \bibinfo {author} {\bibfnamefont {H.-T.}\ \bibnamefont {Jeng}},
  \bibinfo {author} {\bibfnamefont {S.-K.}\ \bibnamefont {Mo}}, \bibinfo
  {author} {\bibfnamefont {Z.}~\bibnamefont {Hussain}}, \bibinfo {author}
  {\bibfnamefont {A.}~\bibnamefont {Bansil}}, \ and\ \bibinfo {author}
  {\bibfnamefont {Z.-X.}\ \bibnamefont {Shen}},\ }\href {\doibase
  10.1038/nnano.2013.277} {\bibfield  {journal} {\bibinfo  {journal} {Nature
  Nanotechnology}\ }\textbf {\bibinfo {volume} {9}},\ \bibinfo {pages} {111}
  (\bibinfo {year} {2014})}\BibitemShut {NoStop}%
\bibitem [{\citenamefont {Marinov}\ \emph {et~al.}(2017)\citenamefont
  {Marinov}, \citenamefont {Avsar}, \citenamefont {Watanabe}, \citenamefont
  {Taniguchi},\ and\ \citenamefont {Kis}}]{marinov_resolving_2017}%
  \BibitemOpen
  \bibfield  {author} {\bibinfo {author} {\bibfnamefont {K.}~\bibnamefont
  {Marinov}}, \bibinfo {author} {\bibfnamefont {A.}~\bibnamefont {Avsar}},
  \bibinfo {author} {\bibfnamefont {K.}~\bibnamefont {Watanabe}}, \bibinfo
  {author} {\bibfnamefont {T.}~\bibnamefont {Taniguchi}}, \ and\ \bibinfo
  {author} {\bibfnamefont {A.}~\bibnamefont {Kis}},\ }\href {\doibase
  10.1038/s41467-017-02047-5} {\bibfield  {journal} {\bibinfo  {journal}
  {Nature Communications}\ }\textbf {\bibinfo {volume} {8}},\ \bibinfo {pages}
  {1938} (\bibinfo {year} {2017})}\BibitemShut {NoStop}%
\bibitem [{\citenamefont {Ko{\'s}mider}\ \emph {et~al.}(2013)\citenamefont
  {Ko{\'s}mider}, \citenamefont {Gonz{\'a}lez},\ and\ \citenamefont
  {Fern{\'a}ndez-Rossier}}]{kosmider_large_2013}%
  \BibitemOpen
  \bibfield  {author} {\bibinfo {author} {\bibfnamefont {K.}~\bibnamefont
  {Ko{\'s}mider}}, \bibinfo {author} {\bibfnamefont {J.~W.}\ \bibnamefont
  {Gonz{\'a}lez}}, \ and\ \bibinfo {author} {\bibfnamefont {J.}~\bibnamefont
  {Fern{\'a}ndez-Rossier}},\ }\href {\doibase 10.1103/PhysRevB.88.245436}
  {\bibfield  {journal} {\bibinfo  {journal} {Physical Review B}\ }\textbf
  {\bibinfo {volume} {88}},\ \bibinfo {pages} {245436} (\bibinfo {year}
  {2013})}\BibitemShut {NoStop}%
\bibitem [{\citenamefont {Yu}\ \emph {et~al.}(2014)\citenamefont {Yu},
  \citenamefont {Wu}, \citenamefont {Liu}, \citenamefont {Xu},\ and\
  \citenamefont {Yao}}]{yu_nonlinear_2014}%
  \BibitemOpen
  \bibfield  {author} {\bibinfo {author} {\bibfnamefont {H.}~\bibnamefont
  {Yu}}, \bibinfo {author} {\bibfnamefont {Y.}~\bibnamefont {Wu}}, \bibinfo
  {author} {\bibfnamefont {G.-B.}\ \bibnamefont {Liu}}, \bibinfo {author}
  {\bibfnamefont {X.}~\bibnamefont {Xu}}, \ and\ \bibinfo {author}
  {\bibfnamefont {W.}~\bibnamefont {Yao}},\ }\href {\doibase
  10.1103/PhysRevLett.113.156603} {\bibfield  {journal} {\bibinfo  {journal}
  {Physical Review Letters}\ }\textbf {\bibinfo {volume} {113}},\ \bibinfo
  {pages} {156603} (\bibinfo {year} {2014})}\BibitemShut {NoStop}%
\bibitem [{\citenamefont {Klein}\ \emph {et~al.}(2016)\citenamefont {Klein},
  \citenamefont {Wierzbowski}, \citenamefont {Regler}, \citenamefont {Becker},
  \citenamefont {Heimbach}, \citenamefont {M{\"u}ller}, \citenamefont
  {Kaniber},\ and\ \citenamefont {Finley}}]{klein_stark_2016}%
  \BibitemOpen
  \bibfield  {author} {\bibinfo {author} {\bibfnamefont {J.}~\bibnamefont
  {Klein}}, \bibinfo {author} {\bibfnamefont {J.}~\bibnamefont {Wierzbowski}},
  \bibinfo {author} {\bibfnamefont {A.}~\bibnamefont {Regler}}, \bibinfo
  {author} {\bibfnamefont {J.}~\bibnamefont {Becker}}, \bibinfo {author}
  {\bibfnamefont {F.}~\bibnamefont {Heimbach}}, \bibinfo {author}
  {\bibfnamefont {K.}~\bibnamefont {M{\"u}ller}}, \bibinfo {author}
  {\bibfnamefont {M.}~\bibnamefont {Kaniber}}, \ and\ \bibinfo {author}
  {\bibfnamefont {J.~J.}\ \bibnamefont {Finley}},\ }\href {\doibase
  10.1021/acs.nanolett.5b03954} {\bibfield  {journal} {\bibinfo  {journal}
  {Nano Letters}\ }\textbf {\bibinfo {volume} {16}},\ \bibinfo {pages} {1554}
  (\bibinfo {year} {2016})}\BibitemShut {NoStop}%
\bibitem [{\citenamefont {Chu}\ \emph {et~al.}(2015)\citenamefont {Chu},
  \citenamefont {Ilatikhameneh}, \citenamefont {Klimeck}, \citenamefont
  {Rahman},\ and\ \citenamefont {Chen}}]{chu_electrically_2015}%
  \BibitemOpen
  \bibfield  {author} {\bibinfo {author} {\bibfnamefont {T.}~\bibnamefont
  {Chu}}, \bibinfo {author} {\bibfnamefont {H.}~\bibnamefont {Ilatikhameneh}},
  \bibinfo {author} {\bibfnamefont {G.}~\bibnamefont {Klimeck}}, \bibinfo
  {author} {\bibfnamefont {R.}~\bibnamefont {Rahman}}, \ and\ \bibinfo {author}
  {\bibfnamefont {Z.}~\bibnamefont {Chen}},\ }\href {\doibase
  10.1021/acs.nanolett.5b03218} {\bibfield  {journal} {\bibinfo  {journal}
  {Nano Letters}\ }\textbf {\bibinfo {volume} {15}},\ \bibinfo {pages} {8000}
  (\bibinfo {year} {2015})}\BibitemShut {NoStop}%
\bibitem [{\citenamefont {Polizzi}(2009)}]{PhysRevB.79.115112}%
  \BibitemOpen
  \bibfield  {author} {\bibinfo {author} {\bibfnamefont {E.}~\bibnamefont
  {Polizzi}},\ }\href {\doibase 10.1103/PhysRevB.79.115112} {\bibfield
  {journal} {\bibinfo  {journal} {Phys. Rev. B}\ }\textbf {\bibinfo {volume}
  {79}},\ \bibinfo {pages} {115112} (\bibinfo {year} {2009})}\BibitemShut
  {NoStop}%
\bibitem [{\citenamefont {Balay}\ \emph {et~al.}(2019)\citenamefont {Balay},
  \citenamefont {Abhyankar}, \citenamefont {Adams}, \citenamefont {Brown},
  \citenamefont {Brune}, \citenamefont {Buschelman}, \citenamefont {Dalcin},
  \citenamefont {Dener}, \citenamefont {Eijkhout}, \citenamefont {Gropp},
  \citenamefont {Karpeyev}, \citenamefont {Kaushik}, \citenamefont {Knepley},
  \citenamefont {May}, \citenamefont {McInnes}, \citenamefont {Mills},
  \citenamefont {Munson}, \citenamefont {Rupp}, \citenamefont {Sanan},
  \citenamefont {Smith}, \citenamefont {Zampini}, \citenamefont {Zhang},\ and\
  \citenamefont {Zhang}}]{petsc-web-page}%
  \BibitemOpen
  \bibfield  {author} {\bibinfo {author} {\bibfnamefont {S.}~\bibnamefont
  {Balay}}, \bibinfo {author} {\bibfnamefont {S.}~\bibnamefont {Abhyankar}},
  \bibinfo {author} {\bibfnamefont {M.~F.}\ \bibnamefont {Adams}}, \bibinfo
  {author} {\bibfnamefont {J.}~\bibnamefont {Brown}}, \bibinfo {author}
  {\bibfnamefont {P.}~\bibnamefont {Brune}}, \bibinfo {author} {\bibfnamefont
  {K.}~\bibnamefont {Buschelman}}, \bibinfo {author} {\bibfnamefont
  {L.}~\bibnamefont {Dalcin}}, \bibinfo {author} {\bibfnamefont
  {A.}~\bibnamefont {Dener}}, \bibinfo {author} {\bibfnamefont
  {V.}~\bibnamefont {Eijkhout}}, \bibinfo {author} {\bibfnamefont {W.~D.}\
  \bibnamefont {Gropp}}, \bibinfo {author} {\bibfnamefont {D.}~\bibnamefont
  {Karpeyev}}, \bibinfo {author} {\bibfnamefont {D.}~\bibnamefont {Kaushik}},
  \bibinfo {author} {\bibfnamefont {M.~G.}\ \bibnamefont {Knepley}}, \bibinfo
  {author} {\bibfnamefont {D.~A.}\ \bibnamefont {May}}, \bibinfo {author}
  {\bibfnamefont {L.~C.}\ \bibnamefont {McInnes}}, \bibinfo {author}
  {\bibfnamefont {R.~T.}\ \bibnamefont {Mills}}, \bibinfo {author}
  {\bibfnamefont {T.}~\bibnamefont {Munson}}, \bibinfo {author} {\bibfnamefont
  {K.}~\bibnamefont {Rupp}}, \bibinfo {author} {\bibfnamefont {P.}~\bibnamefont
  {Sanan}}, \bibinfo {author} {\bibfnamefont {B.~F.}\ \bibnamefont {Smith}},
  \bibinfo {author} {\bibfnamefont {S.}~\bibnamefont {Zampini}}, \bibinfo
  {author} {\bibfnamefont {H.}~\bibnamefont {Zhang}}, \ and\ \bibinfo {author}
  {\bibfnamefont {H.}~\bibnamefont {Zhang}},\ }\href
  {https://www.mcs.anl.gov/petsc} {\enquote {\bibinfo {title} {{PETS}c {W}eb
  page},}\ }\bibinfo {howpublished} {\url{https://www.mcs.anl.gov/petsc}}
  (\bibinfo {year} {2019})\BibitemShut {NoStop}%
\bibitem [{\citenamefont {Trushin}\ \emph {et~al.}(2018)\citenamefont
  {Trushin}, \citenamefont {Goerbig},\ and\ \citenamefont
  {Belzig}}]{trushin_model_2018}%
  \BibitemOpen
  \bibfield  {author} {\bibinfo {author} {\bibfnamefont {M.}~\bibnamefont
  {Trushin}}, \bibinfo {author} {\bibfnamefont {M.~O.}\ \bibnamefont
  {Goerbig}}, \ and\ \bibinfo {author} {\bibfnamefont {W.}~\bibnamefont
  {Belzig}},\ }\href {\doibase 10.1103/PhysRevLett.120.187401} {\bibfield
  {journal} {\bibinfo  {journal} {Physical Review Letters}\ }\textbf {\bibinfo
  {volume} {120}},\ \bibinfo {pages} {187401} (\bibinfo {year}
  {2018})}\BibitemShut {NoStop}%
\bibitem [{\citenamefont {Bieniek}\ \emph {et~al.}(2019)\citenamefont
  {Bieniek}, \citenamefont {Szulakowska},\ and\ \citenamefont
  {Hawrylak}}]{bieniek_to_be_published}%
  \BibitemOpen
  \bibfield  {author} {\bibinfo {author} {\bibfnamefont {M.}~\bibnamefont
  {Bieniek}}, \bibinfo {author} {\bibfnamefont {L.}~\bibnamefont
  {Szulakowska}}, \ and\ \bibinfo {author} {\bibfnamefont {P.}~\bibnamefont
  {Hawrylak}},\ }\href@noop {} {\  (\bibinfo {year} {2019})},\ \bibinfo {note}
  {to be published}\BibitemShut {NoStop}%
\end{thebibliography}%

\end{document}